\documentclass[aps,prd,twocolumn,showpacs,nofootinbib]{revtex4-1}

\usepackage{graphicx}
\usepackage{amsmath}
\usepackage{bm}
\usepackage{slashed}
\usepackage{epsfig}
\usepackage{amsfonts}
\usepackage{epstopdf}
\usepackage{color}
\usepackage{extarrows}
\usepackage{multirow}
\usepackage[utf8]{inputenc}
\usepackage{hyperref}

\allowdisplaybreaks

\begin{document}

\title{Form Factors for $B_c^*\to \eta_c+l\nu_l$ at NLO in QCD}

\author{Wei Tao$^{a}$}
\email{taowei@htu.edu.cn}
\author{Ya-Hui Zhao$^{a}$}
\email{2402183096@stu.htu.edu.cn}
\author{Li-Ting Wang$^{a}$}
\email{wanglt@htu.edu.cn}
\author{Qin Chang$^{a}$}
\email{changqin@htu.edu.cn}
\author{Zhen-Jun Xiao$^{b}$}
\email{xiaozhenjun@njnu.edu.cn}

\affiliation{$^a$ Institute of Particle and Nuclear Physics,
	Henan Normal University, Xinxiang 453007, China\\
	$^b$ Department of Physics and Institute of Theoretical Physics, Nanjing Normal University, Nanjing, Jiangsu 210023, China}

\date{\today}

\begin{abstract}

We present the Non-Relativistic QCD (NRQCD) calculations at the next-to-leading order (NLO) of $\alpha_s$ for $B_c^*\to \eta_c$ vector, axial-vector, tensor and axial-tensor form factors, and  obtain complete analytical expressions for the form factors, along with their asymptotic forms in the hierarchical heavy quark limit. Our results show that the NLO corrections are both sizable and well-behaved in the low squared transfer momentum $(q^2)$ region. Using the NRQCD + lattice + $z$-series method, we further provide theoretical predictions for $B_c^*\to \eta_c$ form factors  across the full physical $q^2$ range.
Based on these predicted form factors, we finally compute the decay widths and branching fractions for the semileptonic decays $B_c^*\to \eta_c+l{\nu_l}$.

\end{abstract}

\maketitle

\section{Introduction}

The $B_c^*$ meson, as the only known vector meson composed of two heavy quarks with distinct flavors ($b$ and $c$), plays a unique role in probing the dynamics of strong and electroweak interactions within the Standard Model (SM). Unlike its pseudoscalar counterpart $B_c$, which was experimentally discovered decades ago through semileptonic decays like $B_c\to J/\psi+l {\nu}_l$~\cite{CDF:1998ihx}, the ground-state vector $B_c^*$ meson remains elusive in experiments~\cite{CMS:2019uhm,LHCb:2019bem}, owing to
its low production rate and detection efficiency.
The ground-state $B_c^*$ meson only decays through electromagnetic and weak interactions. The dominant decay mode of $B_c^*$ is the electromagnetic transition $B_c^*\to B_c\gamma$,  where the emitted soft photon is difficult to detect  in the hadronic environment of current experiments~\cite{CMS:2019uhm,LHCb:2019bem}. Therefore, studying the weak decays of $B_c^*$, such as the semileptonic decays $B_c^*\to J/\psi(\eta_c)+l {\nu}_l$ can provide a good opportunity and an important and useful alternative to help search for $B_c^*$.

Compared to  $B_c^* \to J/\psi + l \nu_l $, the semileptonic process  $B_c^* \to \eta_c + l \nu_l$  may be messier experimentally,
as the vector meson $J/\psi$ has clean leptonic decays and a narrow mass peak, while the pseudoscalar meson $\eta_c$ decays hadronically, with a large width and significant combinatorial backgrounds~\cite{ParticleDataGroup:2024cfk, Workman:2022ynf}.
Nevertheless, the  $B_c^* \to \eta_c + l \nu_l$  decay is particularly significant because it features a theoretically cleaner background,
in the sense that it involves fewer polarization vectors and form factors, and reduces the complexity of analyzing angular distributions or helicity amplitudes,
thereby simplifying and reducing theoretical uncertainties.

The  $B_c^* \to \eta_c$  transition form factors play a crucial role in encapsulating both the perturbative and non-perturbative QCD effects,
making them essential physical quantities for calculating the decay widths and branching fractions of the semileptonic decays $B_c^*\to \eta_c+l \nu_l$.
Current theoretical studies of $B_c^*$ decays predominantly rely on phenomenological models such as the light-front quark model (LFQM)~\cite{Chang:2018mva,Yang:2022jqu,Wang:2024cyi} and the Bauer-Stech-Wirbel (BSW) model~\cite{Chang:2018mva,R:2019uyb}, as well as nonperturbative methods like QCD sum rules (QCDSR)~\cite{Wang:2012hu}, which have systematically evaluated $B_c^*\to\eta_c$ vector and axial-vector form factors. However, these models and methods often lack rigorous perturbative QCD corrections, leading to significant uncertainties in predictions. While lattice QCD has achieved remarkable progress in calculating vector, axial-vector, tensor and axial-tensor form factors for $B_c\to J/\psi(\eta_c)$ transitions~\cite{Colquhoun:2016osw,Harrison:2020gvo,Harrison:2025yan}, analogous results for $B_c^*$ decays remain absent.

The NRQCD effective theory offers a systematic approach to calculating the higher-order perturbative corrections to these form factors.
It factorizes the physical quantities into short-distance coefficients (calculable via perturbation theory) and long-distance matrix elements (non-perturbative), valid when the heavy quark relative velocity $v\ll 1$, applicable to $B_c^{(*)}$ systems~\cite{Bodwin:1994jh}.
For the pseudoscalar $B_c$ meson, the NLO perturbative QCD corrections to the vector, axial-vector, tensor and axial-tensor form factors for semileptonic decays into $J/\psi$ or $\eta_c$ have been explored, showing significant impacts on decay widths and branching fractions~\cite{Bell:2005gw,Bell:2006tz,Qiao:2011yz,Qiao:2012vt,Tao:2022yur}.
For the vector $B_c^*$ meson, Refs.~\cite{Geng:2023ffc,Chang:2025rna} have calculated NLO QCD corrections to the vector, axial-vector, tensor and axial-tensor form factors for the $B_c^*\to J/\psi$ transition.
In this work, we aim to fill the missing piece  by 
performing the analogous NLO calculations for 
the $B_c^*\to \eta_c$ transition.
Performing higher-order perturbative calculations for the form factors not only allows one to test the convergence of the perturbative expansion in $\alpha_s$ and the dependence on the renormalization scale $\mu$ within the NRQCD factorization framework, but also enables more precise theoretical predictions and further tests of the SM. Moreover, the calculation of (axial-)tensor form factors can provide valuable insights into potential new physics beyond the SM.

The structure of the paper is as follows.
In Sec.~\ref{DEFINITION}, we define the (axial-)vector and (axial-)tensor form factors relevant to the $B_c^*\to \eta_c$ transition.
Sec.~\ref{CALCULATIONPROCESS} outlines the calculation steps for these form factors at NLO within NRQCD.
Analytical NLO results, along with their asymptotic expressions in the hierarchical heavy quark limit, are collected in Sec.~\ref{ANALYTICALRESULTS} and  the Appendix~\ref{appendix123}.
Then Sec.~\ref{NUMERICALRESULTS} provides a detailed numerical analysis at NLO and the NRQCD + lattice + $z$-series theoretical prediction for the form factors.
The phenomenological applications of form factors to decay widths and branching fractions are  presented in Sec.~\ref{PHENOMENOLOGICAL}.
Finally, a summary is given in Sec.~\ref{SUMMARY}.

\section{Definition of Form Factors}\label{DEFINITION}

The vector, axial-vector, tensor, and axial-tensor form factors for the transition $B_c^*\to\eta_c$ are defined via the hadronic transition matrix elements of the corresponding quark currents.
For the vector and axial-vector currents, the matrix elements contain structures involving the polarization vector $\epsilon$, the transfer momentum $q=p-p'$, and the total momentum $P=p+p'$, and can be decomposed into four independent form factors $V$ and $A_{0,1,2}$~\cite{Wang:2024cyi,Yang:2022jqu,R:2019uyb,Chang:2018mva,Wirbel:1985ji,Wang:2018ryc,Wang:2016dkd,Shen:2008zzb,Wang:2007ys,Chang:2018sud,Wang:2012hu} as follows:
\begin{align}
&\left\langle \eta_c\left( p^{\prime }\right)\left|\bar{b} \gamma_\mu c\right| B_c^*\left(\epsilon, p\right)\right\rangle
=
\frac{i \varepsilon_{\mu \nu \alpha \beta} \epsilon^{\nu}P^\alpha q^\beta}{m_{B_c^*}+m_{\eta_c}}V,
\\
&\left\langle \eta_c\left( p^{\prime }\right)\left|\bar{b} \gamma_\mu \gamma_5 c\right| B_c^*\left(\epsilon, p\right)\right\rangle\nonumber\\
=&~\frac{ 2 m_{B_c^*}\epsilon \cdot q q_\mu}{q^2}A_0
+ \left(\epsilon_\mu-\frac{\epsilon \cdot q}{q^2} q_\mu\right) (m_{B_c^*}+m_{\eta_c})A_1
\nonumber\\&
+ \epsilon \cdot q \left(\frac{P_\mu}{m_{B_c^*}+m_{\eta_c}}-\frac{m_{B_c^*}-m_{\eta_c}}{q^2} q_\mu\right) A_2
,
\end{align}
where $m_{B_c^*(\eta_c)}$ is the ${B_c^*(\eta_c)}$ mass, and $q^2$ satisfies the physical constraint $0\leq q^2\leq (m_{B_c^*}-m_{\eta_c})^2$.

Similarly, for the tensor and axial-tensor currents, the corresponding matrix elements can be decomposed into three independent form factors $T$ and $T'_{1,2}$ as follows:
\begin{align}
&\left\langle \eta_c\left( p^{\prime }\right)\left|\bar{b} \sigma_{\mu\nu} q^\nu c\right| B_c^*\left(\epsilon, p\right)\right\rangle
= -\varepsilon_{\mu \nu \alpha \beta} \epsilon^{\nu}P^\alpha q^\beta T,\\
&\left\langle\eta_c\left( p^{\prime }\right)\left|\bar{b}  \sigma_{\mu\nu}\gamma_5 q^\nu  c\right| B_c^*\left(\epsilon, p\right)\right\rangle\nonumber\\
=&~i\left( \epsilon_\mu-\frac{\epsilon\cdot q}{q^2} q_\mu\right) (m_{B_c^*}^2-m_{\eta_c}^2)T'_1\nonumber\\&
-i \epsilon \cdot q \left(P_\mu-\frac{m_{B_c^*}^2-m_{\eta_c}^2}{q^2} q_\mu\right) T'_2
,\label{at120def}
\end{align}
where $\sigma_{\mu\nu}=\frac{i}{2}\left(\gamma_\mu\gamma_\nu-\gamma_\nu\gamma_\mu\right)$.

\section{Computational Framework}\label{CALCULATIONPROCESS}

\begin{figure}[!htbp]
	\centering
	\includegraphics[width=0.45\textwidth]{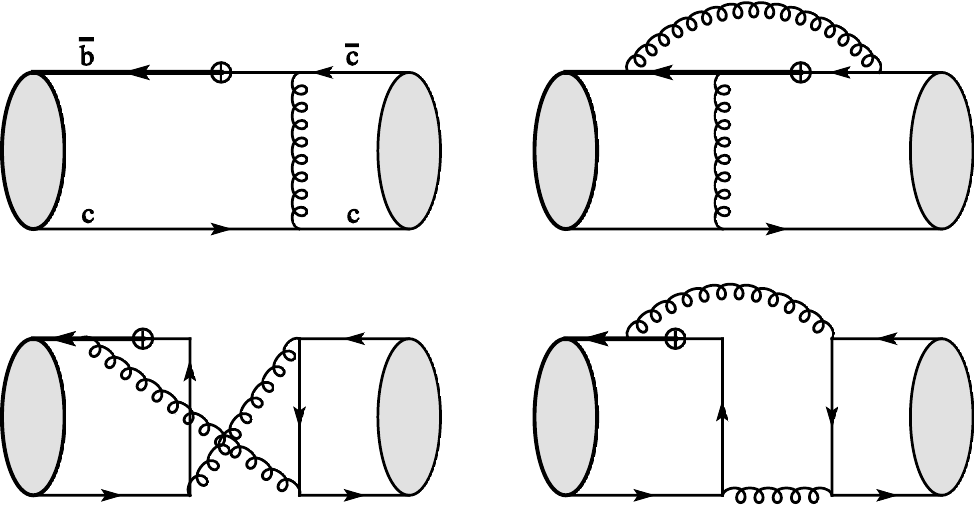}
	\caption{Tree and one-loop sample Feynman diagrams contributing to the form factors, where the circled cross symbol ``$\oplus$'' denotes a certain heavy flavor-changing current vertex. }
	\label{fig:bc2cctree1loop}
\end{figure}
For the calculation of the $B_c^*\to\eta_c$ form factors, we follow the same computational procedure as that used for the $B_c^*\to J/\psi$ form factors in Ref.~\cite{Chang:2025rna}.
As the first step, we employ \texttt{FeynCalc}~\cite{Shtabovenko:2023idz} to generate the Feynman diagrams for the process $c\bar b\to c \bar c W^*$.
Taking into account that the quark loop may involve $n_b$ bottom quarks, $n_c$ charm quarks, and $n_l$ massless quarks, we obtain 2 diagrams at the tree level and 37 diagrams at the one-loop level, with representative examples shown in Fig.~\ref{fig:bc2cctree1loop}.
Note that in Fig.~\ref{fig:bc2cctree1loop}, the sum of the contributions from the two flavor-singlet diagrams in the second row is nonzero.~\footnote{According to the NRQCD calculation, in the semileptonic decay process $B^*_c \to \eta_c$, the total contribution from all six flavor-singlet diagrams to the form factors is nonvanishing and yields a finite complex value, in contrast to the $B_c^*\to J/\psi$ case~\cite{Chang:2025rna}. As a result, the complete NLO result for the $B^*_c \to \eta_c$ form factors contains a small imaginary part, which is approximately $1\%$ of the real part and vanishes in the hierarchical heavy quark limit. For convenience, this small imaginary component will be omitted in the numerical results of the $B^*_c \to \eta_c$ form factors presented in Sec.~\ref{NUMERICALRESULTS} and Sec.~\ref{PHENOMENOLOGICAL}.}

Within the NRQCD factorization framework, the form factors can be decomposed into short-distance coefficients and meson wave functions at the origin~\cite{Qiao:2011yz,Tao:2022yur,Chang:2025rna}. The latter can be extracted with the help of lattice QCD results, while the former originate from the hard contributions of the QCD amplitudes and can be calculated perturbatively to higher orders. To extract the hard QCD amplitudes for the form factors, we employ the covariant projection technique~\cite{Petrelli:1997ge,Bell:2006tz,Jia:2024ini}, and simplify the resulting expressions using the \texttt{FeynCalc}'s functions, \texttt{DiracSimplify}, \texttt{Contract}, \texttt{SUNSimplify}, \texttt{DiracTrace}.

To calculate the trace of a fermion chain containing $\gamma_5$, we adopt the same $\gamma_5$ schemes as described in Ref.~\cite{Chang:2025rna}.
If the fermion chain contains an even number of $\gamma_5$, the naive $\gamma_5$ scheme is applied, in which $\{\gamma_5,\gamma_\mu\}=0$, $\gamma_5^2=1$, and cyclicity of the trace is preserved~\cite{Shtabovenko:2023idz,Tao:2023mtw,Tao:2023vvf,Tao:2023pzv}.
If the fermion chain contains an odd number of $\gamma_5$ and involves the axial-vector or axial-tensor current vertex, $\Gamma_J=\gamma_\mu\gamma_5\text{ or }\sigma_{\mu\nu}\gamma_5$, the fixed reading point $\gamma_5$ scheme is applied~\cite{Shtabovenko:2023idz,Larin:1993tq,Moch:2015usa,Korner:1991sx,Kreimer:1993bh,Kreimer:1989ke,Heller:2020owb,Chen:2022vzo,Sang:2022erv,Li:2023tzx,Vogelsang:1996im,Chen:2024zju,Chen:2023lus}, with the reading point determined by the following replacement rule:
\begin{align}
\text{Trace}\left(a\cdot\Gamma_J\cdot b\right)\to \text{Trace}\left(\frac{\Gamma_J\cdot b\cdot a+b\cdot a\cdot\Gamma_J}{2}\right).
\end{align}
If the fermion chain contains an odd number of $\gamma_5$ and involves neither the axial-vector nor the axial-tensor current vertex, the fixed reading point $\gamma_5$ scheme is applied, with the reading point determined by the following replacement rule:
\begin{align}
\text{Trace}\left(a\cdot\gamma_5\cdot b\right)\to \text{Trace}\left(b\cdot a\cdot\gamma_5\right).
\end{align}

With the aid of the \texttt{TID} function in \texttt{FeynCalc}, the simplified one-loop amplitudes---consisting of scalar products of momenta---can be expressed in terms of the Passarino-Veltman functions $A_0,B_0,C_{0,1},D_0,E_0$. Since the process $c\bar b\to c \bar c W^*$ involves at most three independent propagators, the five-point function $E_0$ can be reduced to lower-point functions $A_0,B_0,C_0$ via integration-by-parts identities~\cite{Chetyrkin:1981qh}. The remaining Passarino-Veltman functions can then be analytically computed with the help of the \texttt{Package-X} software~\cite{Patel:2016fam}. As a result, the form factors can be expressed analytically in terms of elementary and special functions $\texttt{Li}_2$, $\texttt{Log}$, and square roots.  Furthermore, by expanding the form factors in powers of the small charm quark mass $m_c$ and retaining only  the leading-order terms, we obtain their asymptotic expressions in the hierarchical heavy quark limit.

To cancel the divergence appearing in the one-loop diagrams,  we also need to compute the corresponding counterterm diagrams, which consist of  tree-level diagrams with an ${\cal O}(\alpha_s)$ counterterm vertex insertion.  This involves several renormalization procedures: in the on-shell (OS) scheme, the QCD heavy quark field and mass renormalization~\cite{Fael:2020bgs}, as well as the QCD heavy flavor-changing current renormalization; and in the modified-minimal-subtraction (${\overline{\mathrm{MS}}}$) scheme, the QCD coupling renormalization~\cite{Mitov:2006xs,Chetyrkin:1997un,vanRitbergen:1997va} and the NRQCD heavy flavor-changing current renormalization (which does not contribute at NLO~\cite{Tao:2022qxa,Tao:2023pzv,Chang:2025rna}).
In particular, the renormalization constants for the QCD heavy flavor-changing vector ($v$), axial-vector $(a)$, tensor $(t)$, and axial-tensor $(t5)$ currents are given by~\cite{Tao:2023vvf,Tao:2023pzv,Chang:2025rna}:
\begin{align}
Z_{v}^\mathrm{{OS}}= Z_{a}^\mathrm{{OS}}&= 1,\\
Z_t^\mathrm{OS}=Z_{t5}^\mathrm{OS}&=1+\frac{\alpha_s C_F}{4\pi}\left(\frac{1}{\epsilon}-\frac{2x \ln x}{1+x}+2\ln y+{\cal O}(\epsilon)\right)\nonumber\\
&~~~~+{\cal O}(\alpha_s^2),
\end{align}
with the dimensionless variables defined as follows:
\begin{align}
x=\frac{m_c}{m_b},~~~~~
y=\frac{\mu}{m_b},~~~~~
s=\frac{1}{1-\frac{q^2}{m_b^2}}.
\end{align}

After renormalization, we find that all divergences cancel, resulting in finite form factors that are  renormalization-group invariant~\cite{Tao:2023mtw,Tao:2023vvf,Tao:2023pzv,Chang:2025rna}, which serves as an important consistency check of our calculation.  In addition, all of our calculations are performed with a general gauge parameter, and the NLO results for the form factors are found to be independent of the gauge parameter, further confirming the correctness of our calculation.

\section{Analytical Results and Asymptotics}\label{ANALYTICALRESULTS}

Within  NRQCD, we express  the leading order (LO) results of  the $B_c^*\to \eta_c$ vector, axial-vector, tensor, and axial-tensor form factors  as follows
\begin{align}\label{completeLO}
&V=\frac{16\sqrt{2} \pi\alpha _s C_F s^2(1+3x)(1+x)^{\frac{3}{2}}\Psi_{B_c^*}(0)\Psi_{\eta_c}(0)}{m_b^3 x^{\frac{3}{2}}(1+s(x-2)x)^2},\nonumber\\
&V=\frac{1+3x}{1+x} A_0=\frac{2s(1+3x)^2}{3+s x(6+7x)}A_1=2A_2,\nonumber\\
&V=\frac{4s^2 (1+3x)^2 (x^2-1)T'_1}{1-s(6+4x-8x^2)-s^2 x(12+10x-8x^2-15x^3)},\nonumber\\
&T'_2=T=\frac{1+s(4+6x+5x^2)}{4s(1+x)^2}A_0,
\end{align}
where $\Psi_{B_c^*(\eta_c)}(0)$ is  the ${B_c^*(\eta_c)}$ wave function at the origin.
In the hierarchical heavy quark limit, the asymptotic forms for the LO form factors read
\begin{align}\label{asympLO}
V&=\frac{16\sqrt{2} \pi\alpha _s C_F s^2\Psi_{B_c^*}(0)\Psi_{\eta_c}(0)}{m_b^3 x^{\frac{3}{2}}}\nonumber\\
&=\frac{4s}{1+4s}T=\frac{4s}{1+4s}T'_2=\frac{4s^2}{6s-1}T'_1\nonumber\\
&=A_0=\frac{2s}{3}A_1=2 A_2.
\end{align}

At NLO, the complete analytical expressions for the form factors are too lengthy to be presented here and are therefore provided in the ancillary file
 \texttt{NRQCDresultsv2.nb}~\cite{SupplementalMaterial} attached to this paper. Instead,  the asymptotic expressions for the NLO-to-LO ratios of the form factors in the hierarchical heavy quark limit  are given in  the Appendix~\ref{appendix123}.

\section{Numerical Results and Theoretical Predictions}\label{NUMERICALRESULTS}

In the numerical evaluation of the form factors, the input parameter values  are set as follows:
\begin{align}
&m_b = 4.75 \, \text{GeV}, ~ m_c = 1.5 \, \text{GeV};\nonumber\\
&n_b = n_c = 1, n_l = 3 \text{ for } \mu \geq m_b;\nonumber\\
&n_b = 0, n_c = 1, n_l = 3 \text{ for } m_c \leq \mu < m_b;\nonumber\\
&n_b = 0, n_c = 0, n_l = 3 \text{ for } \mu < m_c,
\end{align}
and the QCD running coupling  is computed at one-loop accuracy using the \texttt{AsRunDec} function from the \texttt{RunDec} package~\cite{Chetyrkin:2000yt,Schmidt:2012az,Deur:2016tte,Herren:2017osy}.


\begin{figure}[!htbp]
	\centering
	\includegraphics[width=0.45\textwidth]{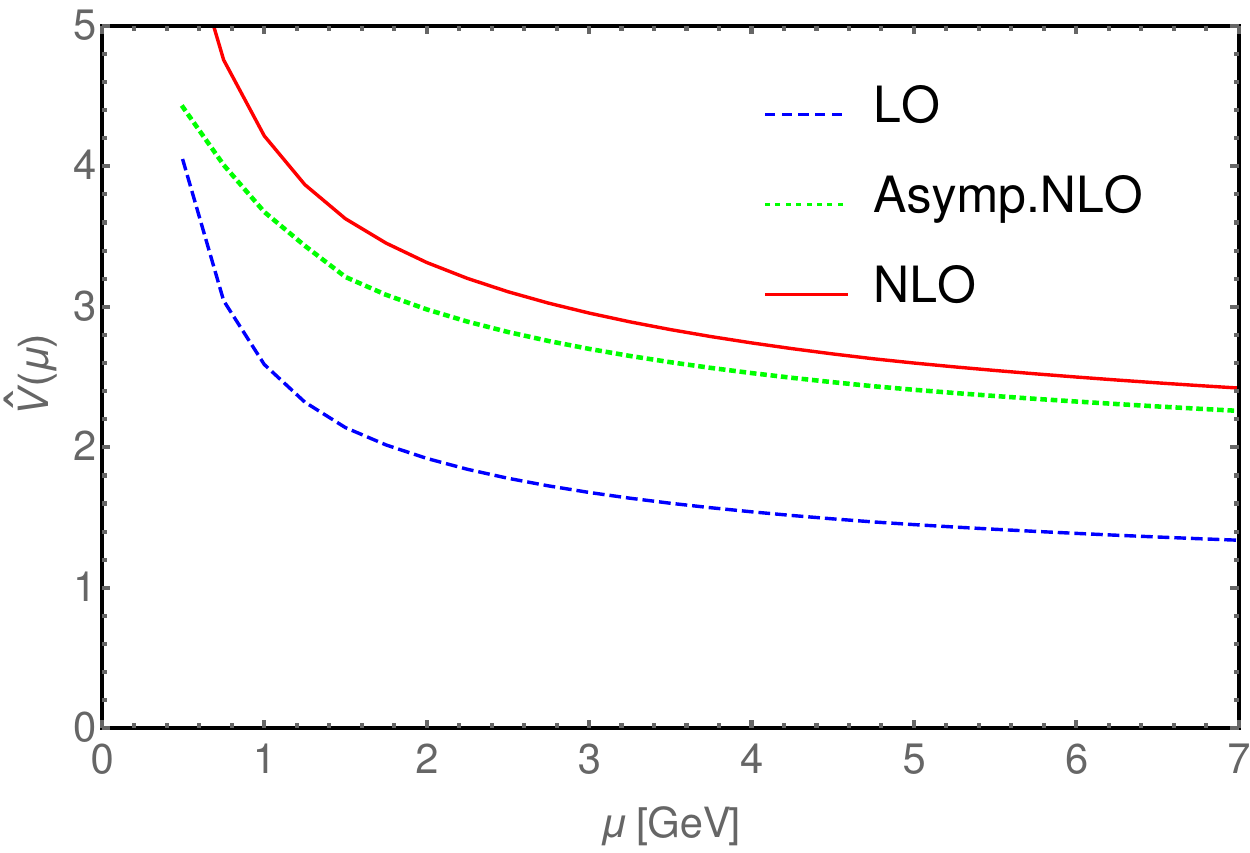}
	\caption{The renormalization scale $\mu$ dependence of the normalized form factor $\hat{V}(\mu)=V(\mu){(m_b/10)^3}/({\Psi_{B_c^*}(0) \Psi_{\eta_c}(0)})$ at LO, asymptotic NLO and complete NLO at the maximum recoil point $q^2=0$.}
	\label{fig:Vmu}
\end{figure}

\begin{figure}[!htbp]
	\centering
	\includegraphics[width=0.45\textwidth]{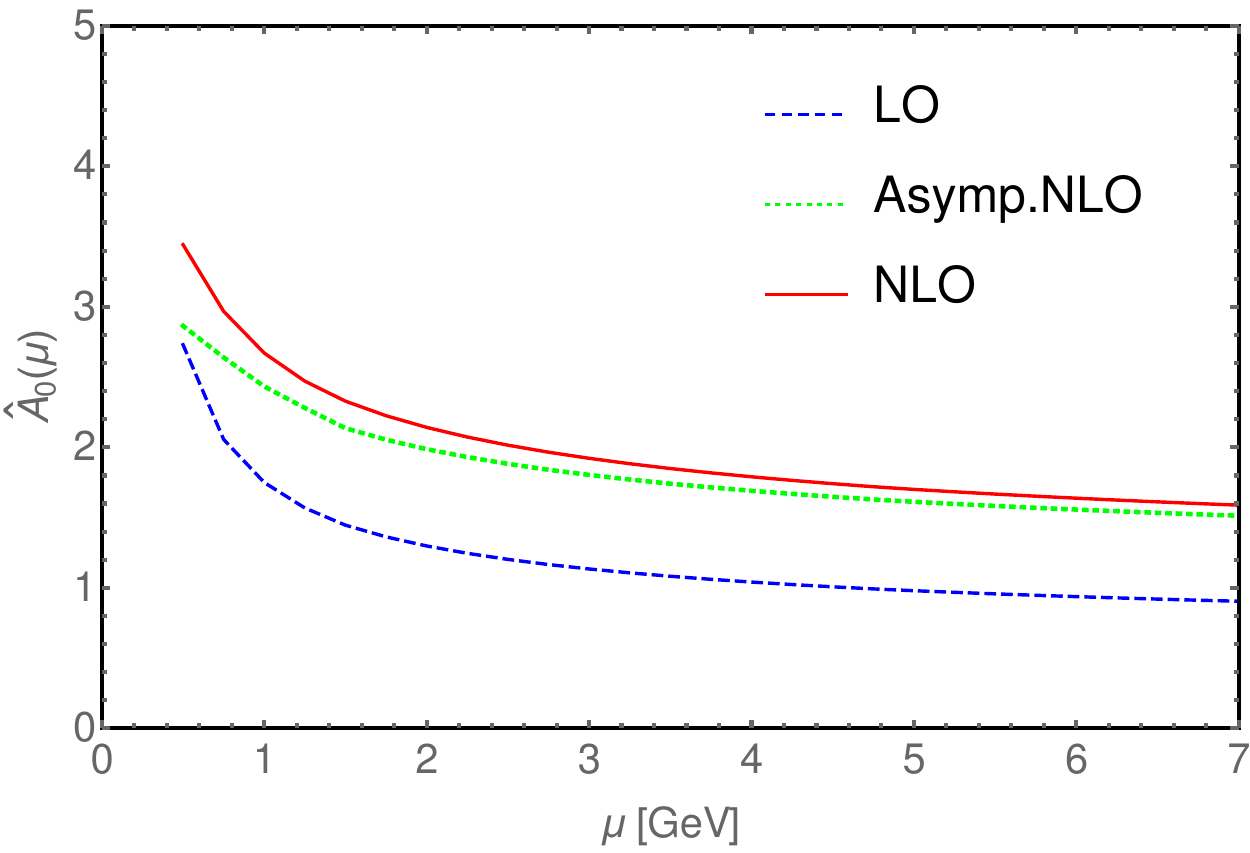}
	\caption{The same as in Fig.~\ref{fig:Vmu}, but for  $\hat{A}_0(\mu)$.}
	\label{fig:A0mu}
\end{figure}

\begin{figure}[!htbp]
	\centering
	\includegraphics[width=0.45\textwidth]{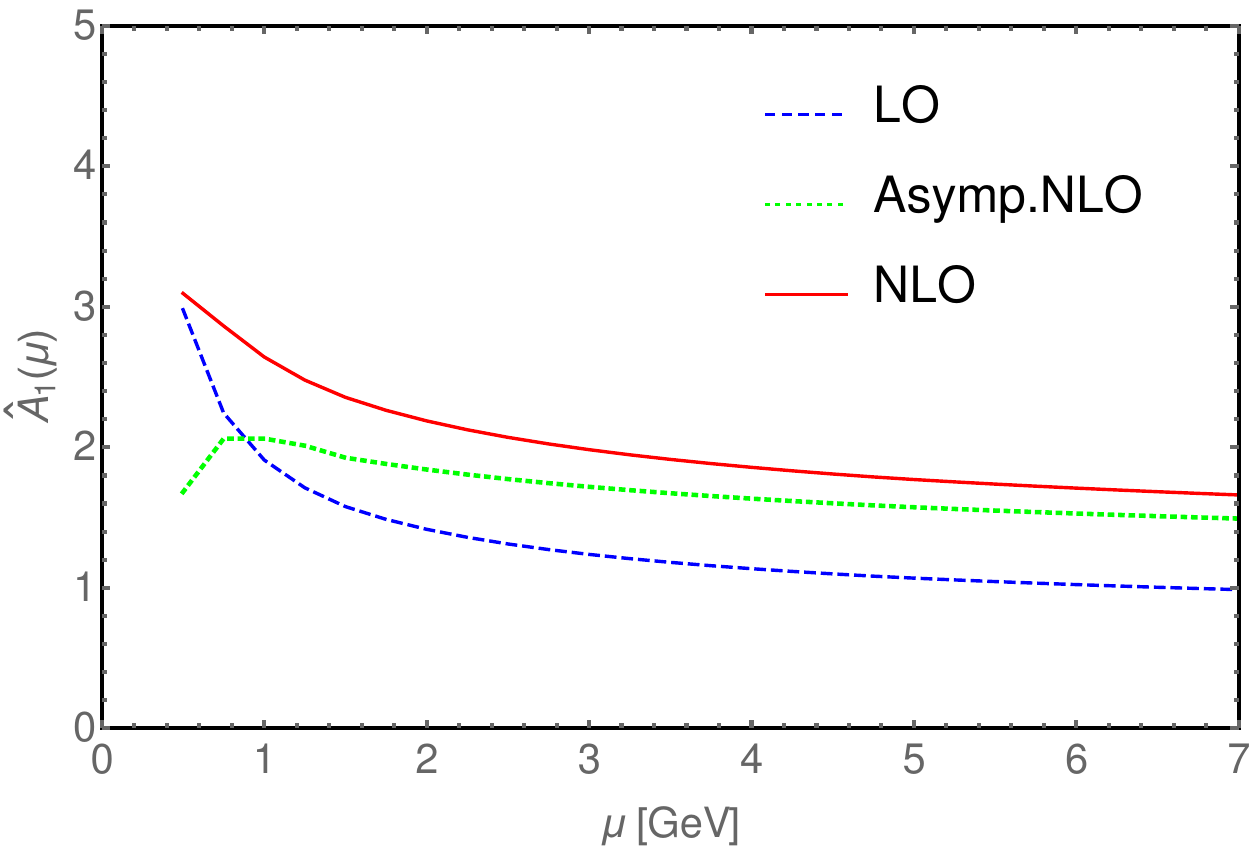}
	\caption{The same as in Fig.~\ref{fig:Vmu}, but for  $\hat{A}_1(\mu)$.}
	\label{fig:A1mu}
\end{figure}

\begin{figure}[!htbp]
	\centering
	\includegraphics[width=0.45\textwidth]{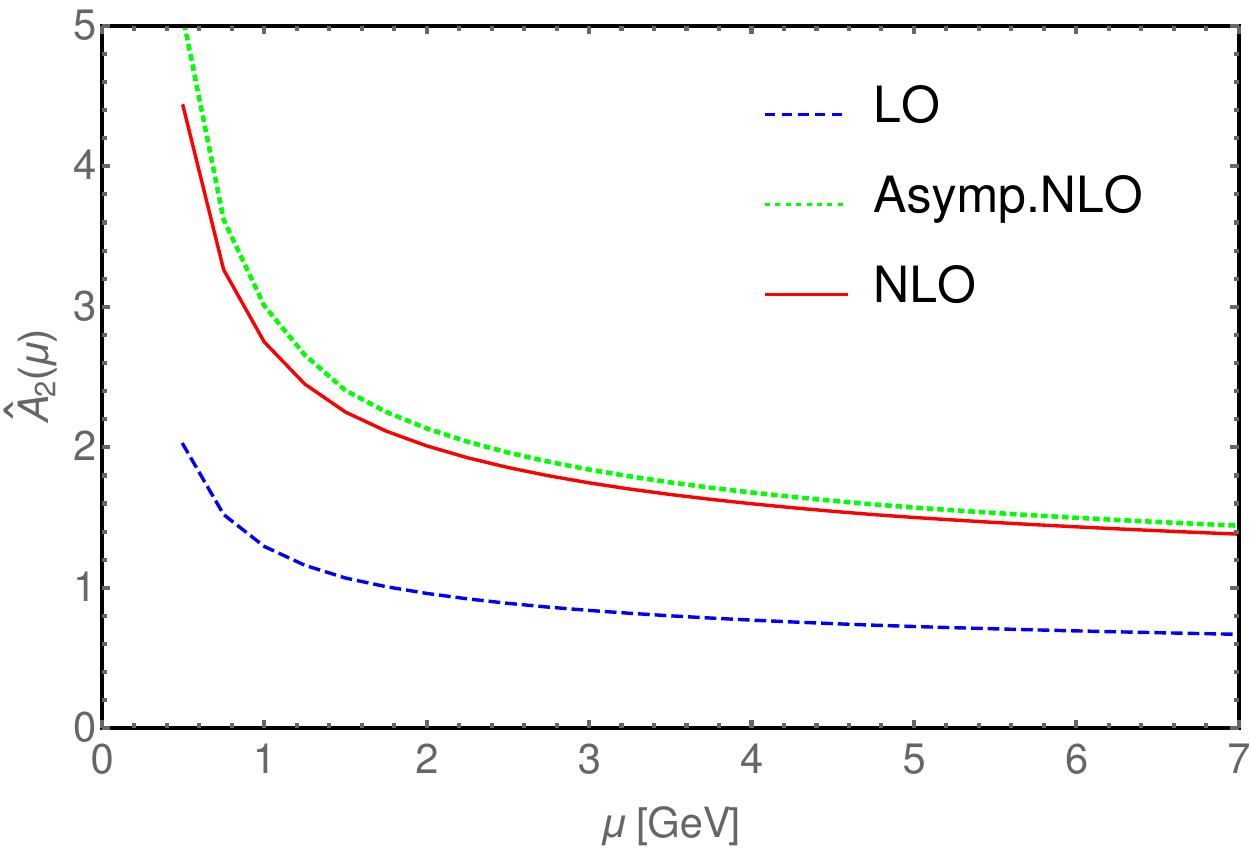}
	\caption{The same as in Fig.~\ref{fig:Vmu}, but for  $\hat{A}_2(\mu)$.}
	\label{fig:A2mu}
\end{figure}

\begin{figure}[!htbp]
	\centering
	\includegraphics[width=0.45\textwidth]{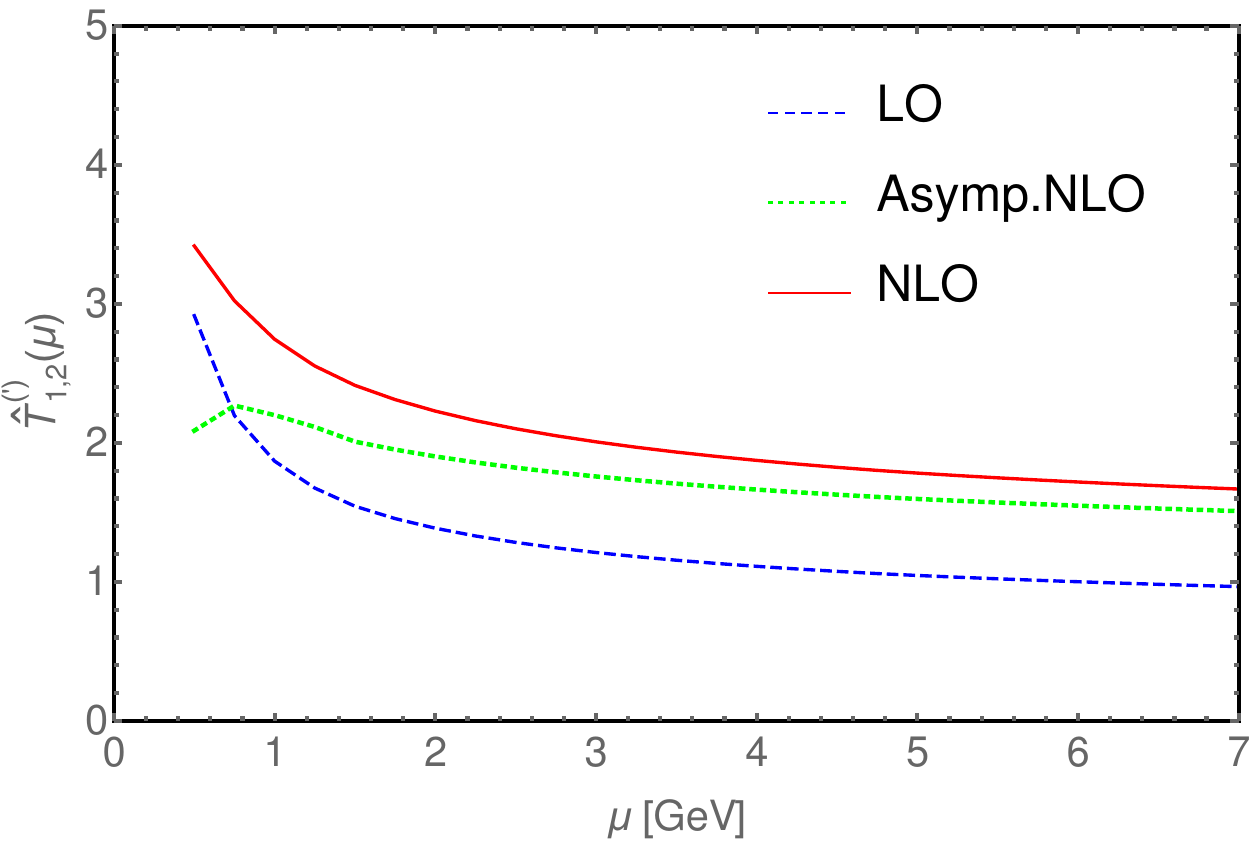}
	\caption{The same as in Fig.~\ref{fig:Vmu}, but for  $\hat{T}(\mu)=\hat{T}'_1(\mu)=\hat{T}'_2(\mu)$.}
	\label{fig:TAT12mu}
\end{figure}

We first investigate the convergence and renormalization scale dependence of the NLO corrections to the form factors.
At the maximum recoil point $(q^2 = 0)$, Figs.~\ref{fig:Vmu}--\ref{fig:TAT12mu} illustrate the renormalization scale $(\mu)$  dependence of the normalized form factor  ${\hat{F}_i(\mu)}={F_i(\mu)}{(m_b/10)^3}/({\Psi_{B_c^*}(0) \Psi_{\eta_c}(0)})$  at LO, asymptotic NLO, and complete NLO accuracies,  where $F_i$ denotes the form factors $V,A_{0,1,2},T,T'_{1,2}$.
As shown in these figures, for most form factors, the NLO corrections are convergent and reduce the renormalization scale dependence. However, an exception is observed for the form factor $A_2$, where the NLO corrections fail to reduce the scale dependence. This behavior is attributed to the presence of excessively large $\alpha_s^2$ terms in the nonconvergent NLO corrections, which will lead to a large $\mu$-dependence at ${\cal O}(\alpha_s^3)$ and offset the renormalization group suppression (see also  Refs.~\cite{Tao:2023mtw,Chang:2025rna} for more details).

As a detail worth mentioning, these curves in Figs.~\ref{fig:Vmu}--\ref{fig:TAT12mu} are plotted through a finite number of sample data points obtained from our NRQCD calculations.
In our numerical calculations, the bottom (charm) quark mass $m_{b(c)}$ has been fixed throughout to its pole mass value, $m_{b(c)} = 4.75 (1.5)~\text{GeV}$,
as we adopt the on-shell quark mass renormalization scheme. Therefore, we do not vary $m_{b(c)}$ with the renormalization scale $\mu$.
By examining the term $(11C_A/3 - 2n_f/3)\ln(2y^{2}/x)$ (which is the only coefficient that depends on $\mu$ in both the asymptotic NLO and full NLO results, due to the renormalization-group invariance) in Eqs.~\eqref{asyeqexp1}--\eqref{asyeqexp5}  of the Appendix~\ref{appendix123}
and using our data provided in the attachment, one can check that as $\mu$ decreases,
the NLO corrections (the  $\alpha_s^2$  term) gradually decrease and change sign from positive to negative at a certain value of  $\mu$,
leading to a peak in the NLO results (the  $\alpha_s$  term +  $\alpha_s^2$  term) in the low  $\mu$  region $0<\mu<1\,\text{GeV}$ (this region may be dominated by non-perturbative effects~\cite{Beneke:1997jm},
which are outside the main focus of the perturbative calculations in this paper, and our discussions regarding this region are only based on our perturbative calculation results).
The peak exists not only in the asymptotic NLO but also in the complete NLO.
Moreover, the peak can be observed not only in the form factors  $A_1$, $T$ and  $T'_{1,2}$,
but also in the other form factors.
However, the values of  $\mu$   where the peaks occur  may differ, and some peaks are only visible at  $\mu$  values even smaller than $0.5\,\text{GeV}$,
which are not plotted in the figures.

It can be seen from the figures that the value of $\mu$ corresponding to the peak of the asymptotic NLO results for $A_1$, $T$, and $T'_{1,2}$ appears to be larger both than that of their corresponding full NLO results and than that of the other form factors. The former discrepancy may be attributed to the absence of higher-order terms in $x$ (one can check this using our data provided in the attachment) in the asymptotic NLO expressions, where only the leading term in the power expansion in $x$ is retained in the hierarchical heavy quark limit.
The latter difference is likely because the positive terms in the coefficients of the $\alpha_s$ corrections  in the asymptotic NLO results of $A_1$, $T$, and $T'_{1,2}$ are relatively smaller compared to those of $V$ and $A_0$ (specifically, comparing the term $\frac{C_F}{4\pi}\frac{7\pi^2}{9}$ against $\frac{C_F}{4\pi}\frac{5\pi^2}{3}$), as shown in Eqs.~\eqref{asyeqexp1}--\eqref{asyeqexp5} of the Appendix~\ref{appendix123}. This results in a net negative finite correction at $\mu= 0.5\,\mathrm{GeV}$ (also negative for all $\mu \le 0.5\,\mathrm{GeV}$), 
thereby delaying the point of stability (the peak) to a higher renormalization scale than 0.5 GeV, as observed in the plots. Conversely, for the asymptotic NLO result of $A_2$, the positive terms in the $\alpha_s$ corrections are significantly larger, which leads to the peak likely shifting to a much smaller $\mu$ value compared to the other form factors.

\begin{figure}[!htbp]
	\centering
	\includegraphics[width=0.45\textwidth]{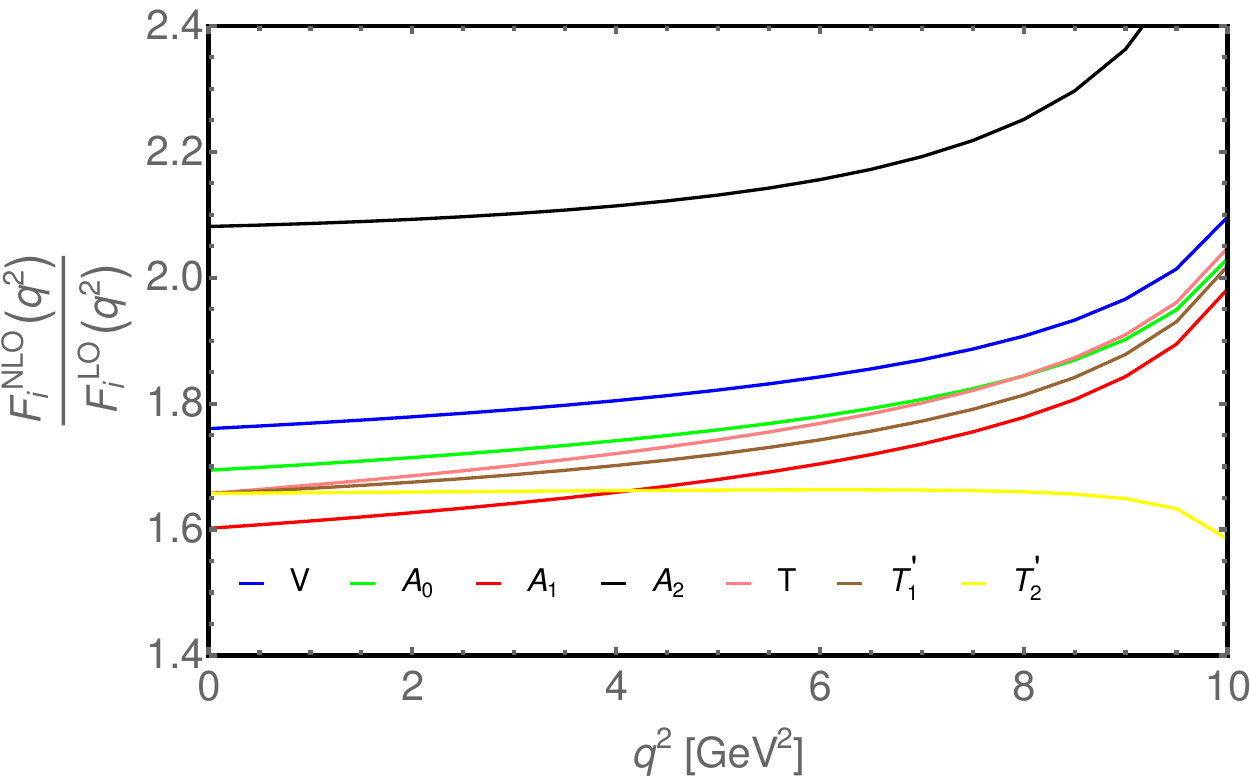}
	\caption{The $q^2$ dependence of $F_i^\text{NLO}(q^2)/F_i^\text{LO}(q^2)$ with  the form factor $F_i$ standing for $V$, $A_{0,1,2}$, $T$ and $T'_{1,2}$. In the computation, the renormalization scale is fixed at $\mu=3\,\text{GeV}$. }
	\label{fig:FiNLO2LOq2}
\end{figure}


To further study how the NLO corrections vary with $q^2$, we compute the NLO-to-LO ratios $F_i^\text{NLO}(q^2)/F_i^\text{LO}(q^2)$  of the form factors (note, throughout the paper, the NLO result $F_i^{\text{NLO}}$ represents 
the form factor evaluated at NLO accuracy and equals
the LO result $F_i^{\text{LO}}$ (the $\alpha_s$ term) plus the NLO correction (the $\alpha_s^2$ term); that is, if $F_i^{\text{LO}} = c_0\,\alpha_s$, then $F_i^{\text{NLO}} = c_0\,\alpha_s + c_1\,\alpha_s^2=F_i^{\text{LO}}(1 + \frac{c_1}{c_0}\,\alpha_s)=F_i^{\text{LO}}(1 +\delta)$) at various values of $q^2$ at the fixed scale $\mu=3\,\text{GeV}$, and plot them in Fig.~\ref{fig:FiNLO2LOq2}. Overall, as can be seen from the figure, the NLO corrections are relatively large ($\delta>50\%$) but remain perturbatively convergent ($\delta<100\%$) in the low $q^2$ region (with the exception of $A_2$, which exhibits particularly large $\mathcal{O}(\alpha_s)$ QCD corrections  ($\delta>100\%$) that are not influenced by higher-order $\mathcal{O}(v^2)$ relativistic corrections). However, in the high $q^2$ region, the  perturbative convergence begins
to break down, suggesting that higher-order NRQCD calculations may not be reliable in the low recoil region~\cite{Qiao:2011yz,Chang:2025rna}.
In addition, the ratio for $T'_2$ appears to be nearly $q^2$-independent up to about $9\, \text{GeV}^2$, with a dip emerging at higher $q^2$ values. 
This behavior is merely a mathematical artifact resulting from our specific choice of the form factor decomposition basis and does not carry particular physical significance. 
To demonstrate this, if
we redefine the form factors, for example as
\begin{align}
&\left\langle\eta_c\left( p^{\prime }\right)\left|\bar{b}  \sigma_{\mu\nu}\gamma_5 q^\nu  c\right| B_c^*\left(\epsilon, p\right)\right\rangle\nonumber\\
=&~i\left(\left( \epsilon_\mu+\frac{\epsilon\cdot q}{q^2} q_\mu\right) (m_{B_c^*}^2-m_{\eta_c}^2)- 2\epsilon \cdot q P_\mu\right)T'_1\nonumber\\&
+i \epsilon \cdot q \left(P_\mu-\frac{m_{B_c^*}^2-m_{\eta_c}^2}{q^2} q_\mu\right) T''_2
,
\end{align}
compared with Eq.~\eqref{at120def}, one can deduce that 
\begin{align}
T_2''=2T_1'-T'_2, 
\end{align}
then it can be found that the ratio for $T_2''$ behaves similarly to the other form factors shown in Fig.~\ref{fig:FiNLO2LOq2}: it increases with $q^2$, and in the high-$q^2$ region the perturbative convergence begins to break down.

To clarify the definitions of the full $q^2$ range and 
the low-/high-$q^2$ regions, 
we make the following remarks. 
In the $\bar{b} \to \bar{x} + l\nu_l (l\bar{l})$ process (note the following conclusions also apply to  the $x\bar{b} \to x\bar{x} + l\nu_l (l\bar{l})$ semi-leptonic decay process),
where $x\in\{u,c\}$ for $l\nu_l$ and $x\in\{d,s\}$ for $l\bar{l}$,
by choosing the center-of-mass frame of  the final lepton pair (either $l\nu_l$ or $l\bar{l}$)~\cite{Richman:1995wm,Cohen:2018vhw},
and letting $Q = (Q_0, \vec{Q})$ and $P = (P_0, \vec{P})$ denote the momenta of the anti-quarks $\bar{b}$ and $\bar{x}$ respectively
(for $x\bar{b} \to x\bar{x} + l\nu_l$, $Q$ and $P$ represent the momenta of the mesons $x\bar{b}$ and $x\bar{x}$, respectively),
and $q = (q_0, \vec{0})$ represent the momentum transferred to the lepton pair, based on momentum conservation,
we have: $q_0 = \sqrt{\vec{Q}^2 + m_b^2} - \sqrt{\vec{Q}^2 + m_x^2}$.
It can be seen that as $\vec{Q}^2 =\vec{P}^2$ increases, $q^2=q_0^2$ decreases.
Based on this, 
one can derive that the full physical range of $q^2$ is given by~\cite{Cooper:2021bkt,Yang:2024jlz,Harrison:2020nrv}: $(m_l^2\,(4m_l^2)\approx 0)\leq q^2\leq ((m_{x\bar b}-m_{x\bar x})^2\approx(m_b+m_x-2m_x)^2)$.
For $x=c$ (i.e. the process $B_c^*\to \eta_c+l\nu_l$), using the values of $m_{B_c^*}$ and $m_{\eta_c}$ in Eqs.~\eqref{paravalues1}--\eqref{paravalues2},
we have $0\leq q^2\leq ((m_{B_c^*}-m_{\eta_c})^2\approx 11.2\,\mathrm{GeV}^2)$.
We also deduce that in the low $q^2$ region (where $q^2=q_0^2$ is small), $\vec{Q}^2 =\vec{P}^2$ is large, meaning that both the $\bar{b}$ and $\bar{c}$ have high energy (carry large momentum),
and the $\bar{b} \to \bar{c} $ transition is perturbative, so the NRQCD factorization is expected to work reliably
and  the NLO NRQCD corrections remain perturbatively convergent with scale-dependence reduced.
While in the high $q^2$ region (where $q^2=q_0^2$ is large), $\vec{Q}^2 =\vec{P}^2$ is small, meaning that both the $\bar{b}$ and $\bar{c}$ are nearly static (at rest) with low energy,
which leads to the $\bar{b} \to \bar{c} $ transition being dominated by the non-perturbative effects
and the perturbative convergence of the NLO expansion in NRQCD beginning to break down.
Thus, according to   Fig.~\ref{fig:FiNLO2LOq2}, depending on whether perturbative convergence is good or poor,
we roughly refer to the region with $0\leq q^2\leq 3\,\mathrm{GeV}^2$ as the low $q^2$ region,
while the region with $7\,\mathrm{GeV}^2\leq q^2\leq 11.2\,\mathrm{GeV}^2$ is considered as the high $q^2$ region.

In order to provide theoretical predictions for the form factors, it is essential to determine the meson wave functions at the origin.
We first approximate the wave function at the origin of the $B_c^*$ meson by that of the $B_c$ meson using heavy quark spin symmetry~\cite{Bodwin:1994jh,Tao:2023pzv},
\begin{align}
\Psi_{B_c^*}(0)\approx\Psi_{B_c}(0).
\end{align}
The product $\Psi_{B_c}(0)\Psi_{\eta_c}(0)$ can be extracted by combining the NLO NRQCD results for the $B_c\to\eta_c$ vector form factors with the corresponding lattice QCD results~\cite{Tao:2022yur,Chang:2025rna,Tang:2022nqm,Colquhoun:2016osw}. Consequently, the form factors for the $B_c^*\to\eta_c$ transition can be approximately evaluated using the following formula:
\begin{align}\label{nrlattformu}
F_{i,\text{NRQCD+lattice}}^{B_c^*\to \eta_c}(q^2)\approx\frac{1}{2}\sum_{j=1}^{2}\frac{F_{i,\text{NRQCD}}^{B_c^*\to \eta_c}(q^2)}{F_{j,\text{NRQCD}}^{B_c\to \eta_c}(q^2)} F_{j,\text{lattice}}^{B_c\to \eta_c}(q^2),
\end{align}
where $F_{j}^{B_c\to \eta_c}$ stands for the ${B_c\to \eta_c}$ vector form factors $f_{+}$ and $f_0$ defined in Refs.~\cite{Qiao:2012vt,Colquhoun:2016osw}.

Note that  Eq.~\eqref{nrlattformu} is evaluated without performing a fixed-order expansion in powers of $\alpha_s$.
Eq.~\eqref{nrlattformu} is based on the standard ``ratio method" (using the NRQCD to predict the ratio of ``vector-to-pseudoscalar'' form factors while using lattice for the normalization). The physical motivation of Eq.~\eqref{nrlattformu} is twofold. 1) Rather than relying solely on the NRQCD determination of the $B_c^*\to\eta_c$ form factors, we use the lattice QCD result for normalization. 2) While the lattice data (Ref.~\cite{Colquhoun:2016osw}) provides the normalization for the pseudoscalar mode ($B_c$), it does not provide the vector mode ($B_c^*$) directly. NRQCD, however, provides a reliable description of the ratio of the form factors, $F_{B_c^*\to\eta_c}/F_{B_c\to\eta_c}$. In this ratio, many systematic uncertainties (such as meson wavefunction uncertainties, renormalization scale dependence and heavy quark mass uncertainties) cancel out.	
The derivation of Eq.~\eqref{nrlattformu} relies on combining heavy quark spin symmetry (HQSS)~\cite{Tao:2023pzv} with the NRQCD factorization formalism.
In NRQCD, the form factor is factorized into a short-distance perturbative coefficient  and the non-perturbative wavefunctions at the origin:
$F_{i,\text{NRQCD}}^{B_c^{(*)}\to \eta_c}(q^2)\approx C_{i,\text{NRQCD}}^{B_c^{(*)}\to \eta_c}(q^2) \cdot \Psi_{B_c^{(*)} }(0) \Psi_{\eta_c}(0)$.
Heavy quark spin symmetry implies that, up to corrections of order $v$ ($v\sim\alpha_s$; see Ref.~\cite{Tao:2023pzv}),
the wave functions at the origin for the pseudoscalar $B_c$ and the vector $B_c^*$ are equal:
$\Psi_{B_c^*}(0) \approx \Psi_{B_c}(0)$.
Note, this equates the spatial probability amplitudes at origin $r=0$ of coordinate space (long-distance non-perturbative intrinsic property), regardless of the decay kinematics at specific values of $q^2$ (short-distance perturbative interaction effects);
in other words, the equality of the quark-antiquark spatial overlap probability within the two mesons is independent of the mesons' external transition kinematics~\cite{Bodwin:1994jh}.
So we can use the approximation $\Psi_{B_c^*}(0)/\Psi_{B_c}(0) \approx 1$
not only at the maximum recoil (where $q^2 = 0$) but also in the low $q^2$ region.
By taking the ratio of the $i$-th $B_c^*$ form factor to the $j$-th $B_c$ form factor in NRQCD at a specific $q^2$,
the unknown non-perturbative wavefunctions cancel out,
leaving a purely perturbative ratio of corresponding short-distance coefficients (calculated via NRQCD at the specific $q^2$),
as described in Ref.~\cite{Tao:2023pzv}:
$\frac{F_{i,\text{NRQCD}}^{B_c^{*}\to \eta_c}(q^2)}{F_{j,\text{NRQCD}}^{B_c\to \eta_c}(q^2)}\approx  \frac{C_{i,\text{NRQCD}}^{B_c^{*}\to \eta_c}(q^2)}{C_{j,\text{NRQCD}}^{B_c\to \eta_c}(q^2)}$.
Multiplying this ratio by the ``true" physical value (at the specific $q^2$) of the $j$-th  $B_c$  form factor, taken from lattice QCD,
then yields the NRQCD + lattice predicted $i$-th  $B_c^*$ form factor  at the specific $q^2$  without needing to explicitly compute the wavefunctions,
i.e. $F_{i,\text{NRQCD+lattice}}^{B_c^*\to \eta_c}(q^2)\approx  \frac{F_{i,\text{NRQCD}}^{B_c^*\to \eta_c}(q^2)}{F_{j,\text{NRQCD}}^{B_c\to \eta_c}(q^2)}\cdot {F_{j,\text{lattice}}^{B_c\to \eta_c}(q^2)}\approx  \frac{C_{i,\text{NRQCD}}^{B_c^*\to \eta_c}(q^2)}{C_{j,\text{NRQCD}}^{B_c\to \eta_c}(q^2)}\cdot {F_{j,\text{lattice}}^{B_c\to \eta_c}(q^2)}$.
It should be noted that the values of $q^2$ in Eq.~\eqref{nrlattformu} cannot be too large, as the NRQCD calculation results are likely valid only in the low $q^2$ region.
Finally, the factor $\frac{1}{2}\sum_{j=1}^2$ in Eq.~\eqref{nrlattformu} is used to average the NRQCD + lattice results derived from the two available lattice inputs ($j=1,2$ corresponding to $f_+$ and $f_0$; see Ref.~\cite{Colquhoun:2016osw}) to reduce errors from choosing a single lattice input.

\begin{table*}[!htbp]
\begin{center}
	\caption{The NRQCD + lattice predictions for the $B_c^*\to \eta_c$ form factors at the maximum recoil point $q^2=0$, where the first uncertainties come from  choosing the renormalization scale as $\mu=3_{-1.5}^{+4}\,\text{GeV}$ in the NRQCD calculation and the second uncertainties come from  the lattice QCD data errors. For comparison, we also list the corresponding LFQM, BSW and QCDSR results. Note that $T^{(')}_{1,2}(0)\equiv T(0)\equiv T'_{1}(0)\equiv T'_{2}(0)$.}
	\label{tab:nrlatt}
	\setlength{\tabcolsep}{1.8mm}
	\renewcommand{\arraystretch}{2.1}
		\begin{tabular}{c|c|c|c|c|c}
			\hline
			\multirow{2}{*}{}
	 		& \multirow{2}{*}{$\text{NRQCD+Lattice}$}
			&  \multicolumn{2}{c|}{$\text{LFQM}$}
            & \multirow{2}{*}{$\text{BSW}$}
            & \multirow{2}{*}{$\text{QCDSR}$}
            \\
            \cline{3-4}
            &
            & \cite{addnote}
            & \cite{Wang:2024cyi}
            & \cite{Chang:2018mva,R:2019uyb}
            & \cite{Wang:2012hu}
   \\
			\hline
			\multirow{1}*{$V(0)$}
			& \multirow{1}*{$0.8062_{+0.0121}^{-0.0082}\pm 0.0682$}
			& \multirow{1}*{$0.91_{-0.02-0.22}^{+0.02+0.18}$}
            & $0.88_{-0.01-0.06}^{+0.01+0.04}$
            & $0.753_{-0.129}^{+0.107}$
            & $0.71\pm 0.14$
              \\
			\hline
			{$	A_0(0) $}
			& {$0.5243_{+0.0012}^{-0.0008}\pm{0.0443}$}
			& $0.66_{-0.01-0.17}^{+0.01+0.14}$
            & $0.56_{-0.02-0.02}^{+0.01+0.01}$
            & $0.526$
            & $0.47\pm 0.09$
	            \\
			\hline
			{$A_1(0)$}
			& {$0.5410_{-0.0095}^{+0.0064}\pm{0.0457}$}
			& $0.69_{-0.01-0.19}^{+0.01+0.17}$
            & $0.59_{-0.02-0.03}^{+0.01+0.02}$
            & $0.561_{-0.096}^{+0.080}$
            & $0.43\pm 0.08$
		                \\
			\hline
			$A_2(0)$
			& {$0.4766_{+0.0314}^{-0.0211}\pm{0.0403}$}
			& $0.59_{-0.02-0.13}^{+0.02+0.12}$
            & $0.45_{-0.02-0.02}^{+0.02+0.01}$
            & $0.495_{-0.062}^{+0.035}$
            & $0.57\pm 0.11$
		               \\
			\hline
			{$T^{(')}_{1,2}(0) $}
			& {$0.5475_{-0.0030}^{+0.0020}\pm{0.0463}$}
			& {$-$}
            & {$-$}
            & {$-$}
            & $-$
		   \\
			\hline
			\end{tabular}
\end{center}		
\end{table*}

\begin{figure}[!htbp]
	\centering
	\includegraphics[width=0.45\textwidth]{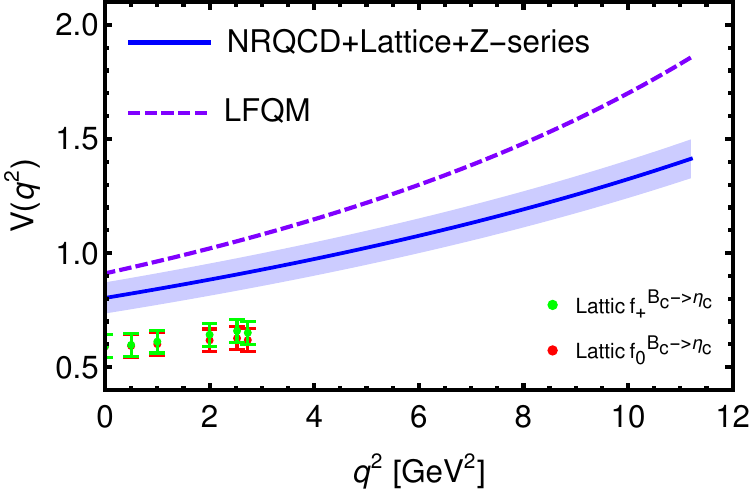}
	\caption{The NRQCD + lattice + $z$-series prediction for the $B_c^*\to \eta_c$ form factor $V(q^2)$ across the full physical $q^2$ range $0\leq q^2\leq (m_{B_c^*}-m_{\eta_c})^2$, where the error band stems from the uncertainties of the input  lattice QCD data.
		The green and red points with error bars denote the  lattice data points with the added $10\%$ uncertainties for the $B_c\to\eta_c$ form factors $f_+$ and $f_0$, respectively, where the central values of the form factors at $q^2\approx  0.00$ and $2.73 \text{ GeV}^2$ are read directly from  Fig.\,2 of Ref.~\cite{Colquhoun:2016osw}, while the central values of the form factors at $q^2 \approx 0.50, \,1.00, \,2.00,$ and $2.52 \text{ GeV}^2$ are estimated from Fig.\,2 of Ref.~\cite{Colquhoun:2016osw}.
		The purple dashed line represents the LFQM prediction from Ref.~\cite{addnote} for 
		the $B_c^*\to \eta_c$ form factor $V(q^2)$. 	
	}
	\label{fig:Vq2}
\end{figure}

\begin{figure}[!htbp]
	\centering
	\includegraphics[width=0.45\textwidth]{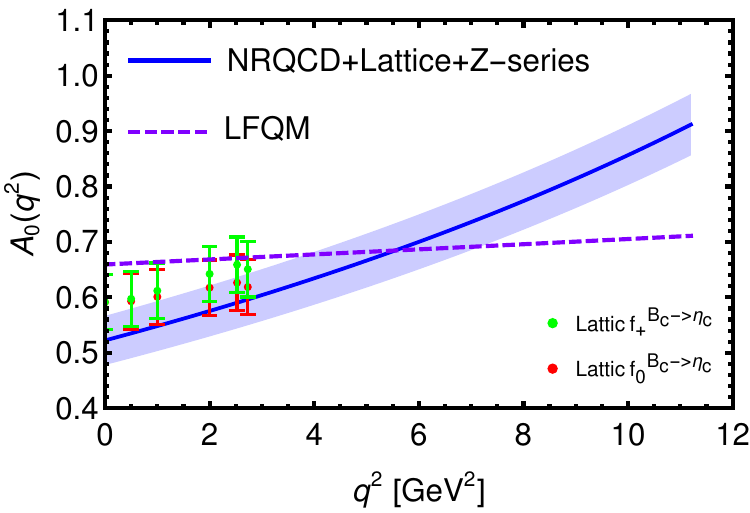}
	\caption{The same as in Fig.~\ref{fig:Vq2}, but for $A_0(q^2)$ of $B_c^*\to \eta_c$.  }
	\label{fig:A0q2}
\end{figure}

\begin{figure}[!htbp]
	\centering
	\includegraphics[width=0.45\textwidth]{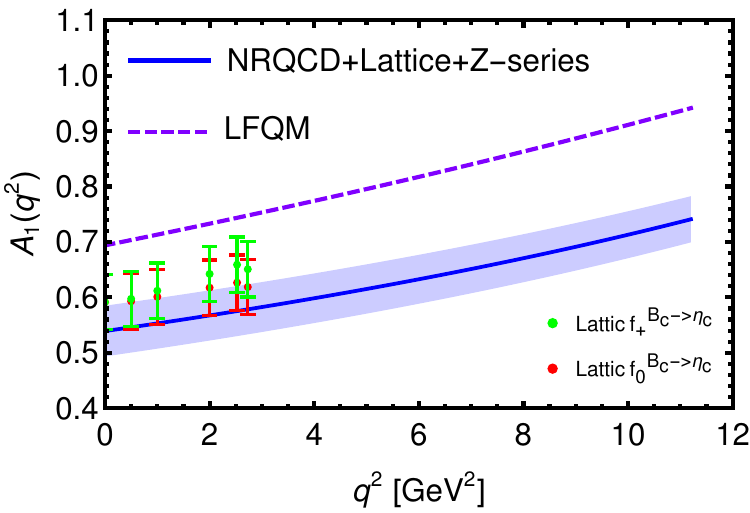}
	\caption{The same as in Fig.~\ref{fig:Vq2}, but for $A_1(q^2)$ of $B_c^*\to \eta_c$.  }
	\label{fig:A1q2}
\end{figure}

\begin{figure}[!htbp]
	\centering
	\includegraphics[width=0.45\textwidth]{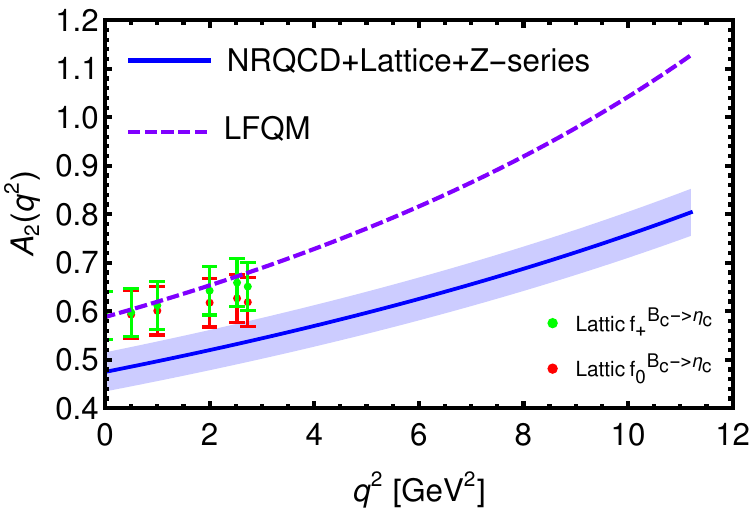}
	\caption{The same as in Fig.~\ref{fig:Vq2}, but for $A_2(q^2)$ of $B_c^*\to \eta_c$.  }
	\label{fig:A2q2}
\end{figure}

\begin{figure}[!htbp]
	\centering
	\includegraphics[width=0.45\textwidth]{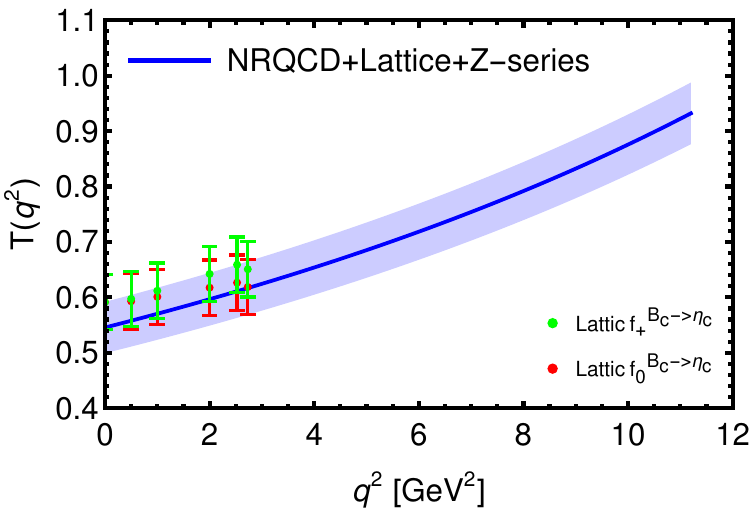}
	\caption{The same as in Fig.~\ref{fig:Vq2}, but for $T(q^2)$ of $B_c^*\to \eta_c$.  }
	\label{fig:Tq2}
\end{figure}

\begin{figure}[!htbp]
	\centering
	\includegraphics[width=0.45\textwidth]{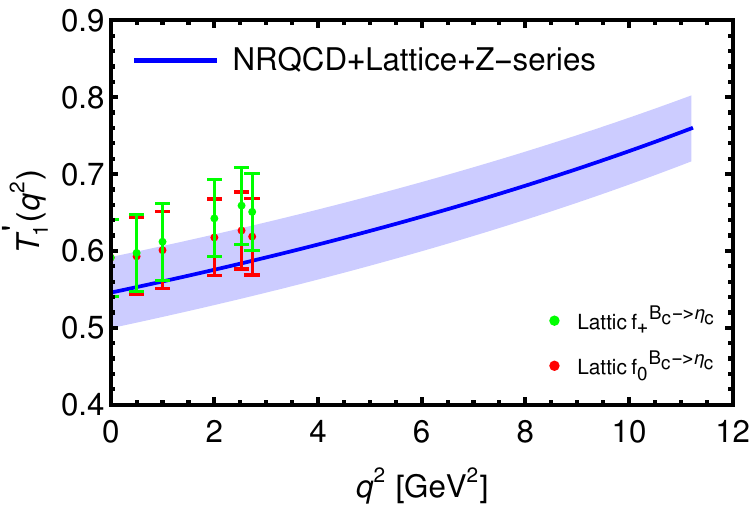}
	\caption{The same as in Fig.~\ref{fig:Vq2}, but for $T'_1(q^2)$ of $B_c^*\to \eta_c$.  }
	\label{fig:AT1q2}
\end{figure}

\begin{figure}[!htbp]
	\centering
	\includegraphics[width=0.45\textwidth]{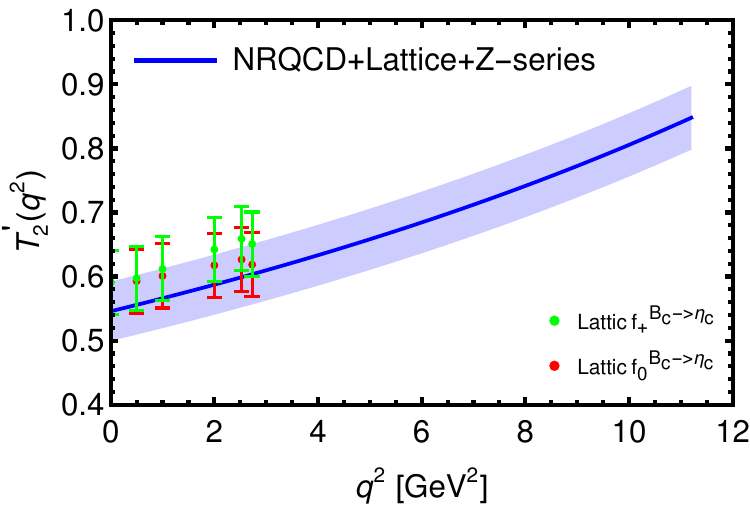}
	\caption{The same as in Fig.~\ref{fig:Vq2}, but for $T'_2(q^2)$ of $B_c^*\to \eta_c$.  }
	\label{fig:AT2q2}
\end{figure}

Regarding the lattice inputs, in the actual calculation of the NRQCD + lattice predictions for the $B_c^* \to \eta_c$ form factors using Eq.~\eqref{nrlattformu},
we have manually added a conservative uncertainty of $\pm 0.05$ ($\approx \pm 10\%$ of original size) to each lattice data point of the $B_c\to\eta_c$ form factors $f_+,f_0$ shown  in the left panel of   Fig. 2 of Ref.~\cite{Colquhoun:2016osw}, and we use the lower and upper bounds of  these uncertainty-augmented data points as the lattice inputs in Eq.~\eqref{nrlattformu}.
Consequently, all subsequent numerical calculations and phenomenological predictions are performed with this added uncertainty taken into account.
Specifically, the uncertainties labeled as coming from lattice QCD data errors in our NRQCD + lattice results in  Table~\ref{tab:nrlatt},
the lower and upper bounds of the NRQCD + lattice form factors  used as input data for the $z$-series fits  in Table~\ref{tab:fit_input}, the corresponding results  of the fitted $z$-series expansion coefficients in Table~\ref{tab:fit_output},
the error bands  of the NRQCD + lattice + $z$-series predictions  shown in  Figs.~\ref{fig:Vq2}--\ref{fig:AT2q2}, and the uncertainties of our calculation results in Table~\ref{tab:widthBr} are all caused by this manually added $\sim 10\%$ uncertainty in the lattice inputs used in Eq.~\eqref{nrlattformu}.

Based on  Eq.~\eqref{nrlattformu}, we first present in Table~\ref{tab:nrlatt} the NRQCD + lattice  predictions for the $B_c^*\to\eta_c$ form factors at $q^2=0$,
with uncertainties arising from the choice of the renormalization scale  $\mu=3_{-1.5}^{+4}\,\text{GeV}$ in the NLO NRQCD calculation and the errors in the lattice QCD data.
The lower limit of the renormalization scale is chosen at the soft (factorization) scale
$\mu_f = \alpha_s(\mu_f) C_F (2 m_b m_c)/(m_b + m_c)$, which is approximately $1.5\,\text{GeV}$ (see Ref.~\cite{Beneke:2014qea}).
The central value is taken to be $\mu_0 =\sqrt{m_b m_c}\approx 3 \,\text{GeV}$,
which is a commonly adopted characteristic hard scale in heavy–quark processes involving both
$b$ and $c$ quarks~\cite{Bell:2006tz, Bell:2005gw, Sang:2022tnh, Tao:2023mtw}.
The upper limit is chosen near the high-energy region
$m_b+m_c\approx 7\,\text{GeV}$, ensuring that the interval
$[3-1.5,3+4]\text{ GeV}$  covers the  perturbative and nonperturbative regions in QCD for double-heavy meson production and decay~\cite{Ai:2025xop,Bell:2006tz,Qiao:2011yz,Egner:2021lxd,Tao:2023mtw}.
In the NRQCD + lattice results in Table~\ref{tab:nrlatt},
the scale uncertainties of the form factors are evaluated by subtracting their central values at $\mu_0=3\text{ GeV}$ from the maximum and minimum values obtained within the range $\mu\in[1.5,7]\text{ GeV}$.
Here, we focus on establishing the central values and the theoretical envelope (defined by the lower and upper limits) of the form factors,
rather than assuming a specific statistical distribution for the scale variation~\cite{Cacciari:2011ze,LHCHiggsCrossSectionWorkingGroup:2011wcg}.

For comparison, the table also includes the corresponding  predictions from the LFQM~\cite{addnote,Wang:2024cyi}, BSW~\cite{Chang:2018mva,R:2019uyb}, and QCDSR~\cite{Wang:2012hu}  methods.
From Table~\ref{tab:nrlatt}, it can be observed that the NRQCD + lattice predictions for the $B_c^*\to\eta_c$ vector and axial-vector form factors agree well with the corresponding predictions from the LFQM, BSW and QCDSR methods. Moreover, in our NRQCD + lattice predictions, the uncertainties are dominated by the errors in the lattice QCD inputs, rather than by the choice of the renormalization scale~\cite{Chang:2025rna}.

To extend our theoretical predictions of the form factors from the low $q^2$ region to the high $q^2$ region, we employ the $z$-series extrapolation method~\cite{Boyd:1997kz,Bourrely:2008za,Hu:2019qcn,Leljak:2019eyw,Harrison:2020gvo,Bharucha:2010im,Hill:2006ub,Biswas:2023bqz}, in which the form factor is parameterized  as a truncated power series in $z(q^2)$,
\begin{align}\label{zseries}
F_{i}(q^2) &=\frac{1}{1-\frac{q^2}{ m_{R}^{2}}} \sum_{n=0}^{N} \alpha_{i,n}\, z^{n}\left(q^2\right),
\end{align}
with
\begin{align}
z(q^2) &=\frac{\sqrt{t_{+}-q^2}-\sqrt{t_{+}-t_{0}}}{\sqrt{t_{+}-q^2}+\sqrt{t_{+}-t_{0}}}.
\end{align}
Here, $t_0 = t_+(1 - \sqrt{1 - t_-/t_+})$ and $t_{\pm} = (m_{B_c^*} \pm m_{\eta_c})^2 \approx (m_b + m_c \pm 2m_c)^2$.
We adopt the quark masses $m_b = 4.75 \text{ GeV}$ and $m_c = 1.5 \text{ GeV}$.
The low-lying $c\bar b$ resonance masses $m_R$ are set to $6.34 \text{ GeV}$ for $V, T$
and $6.75 \text{ GeV}$ for $A_{0,1,2}, T'_{1,2}$~\cite{Boyd:1997kz,Hu:2019qcn,Leljak:2019eyw,Harrison:2025yan}.
Since $z(q^2)$ remains very small throughout the physical $q^2$ range in the process $c\bar b\to c\bar c+l\nu_l$~\cite{Wang:2018duy,Bharucha:2010im,Leljak:2019eyw,Chang:2025rna}, the truncation order is typically taken as $N=1$.
The unknown parameters $\alpha_{i,0}$ and $\alpha_{i,1}$ can then be determined by fitting to the NRQCD + lattice results in the low $q^2$ region.~\footnote{Here, Eq.~\eqref{nrlattformu} (the NRQCD + lattice ratio formula) is used to generate the synthetic data points in the reliable low-$q^2$ region. And Eq.~\eqref{zseries} (the $z$-series expansion) is then used as the fitting function to capture the shape of these points and extrapolate the form factors to the high-$q^2$ region.}
As a result, Eq.~\eqref{zseries} allows us to predict the form factors across the full physical $q^2$ range $0\leq q^2\leq (m_{B_c^*}-m_{\eta_c})^2$.



\begin{table*}[!htbp]
\centering
\caption{\label{tab:fit_input}
Lower bound (LB) and upper bound (UB) values of the NRQCD + lattice results for $B_c^*\to\eta_c$ form factors at different $q^2$ points used as input data for the $z$-series fits. }
\renewcommand{\arraystretch}{1.3} 
\setlength{\tabcolsep}{4pt}       
\resizebox{\textwidth}{!}{
\begin{tabular}{c|c|c|c|c|c|c|c}
\hline
$q^2$ & $V$ & $A_0$ & $A_1$ & $A_2$ & $T$ & $T_1'$ & $T_2'$ \\
$[\text{GeV}^2]$ & (LB/UB) & (LB/UB) & (LB/UB) & (LB/UB) & (LB/UB) & (LB/UB) & (LB/UB) \\
\hline
0.0000 & 0.7380/0.8743 & 0.4799/0.5686 & 0.4953/0.5867 & 0.4363/0.5169 & 0.5012/0.5938 & 0.5012/0.5938 & 0.5012/0.5938 \\ 
0.5000 & 0.7495/0.8869 & 0.4875/0.5770 & 0.4975/0.5888 & 0.4425/0.5237 & 0.5083/0.6015 & 0.5038/0.5963 & 0.5065/0.5993 \\ 
1.0000 & 0.7713/0.9100 & 0.5019/0.5921 & 0.5065/0.5976 & 0.4548/0.5366 & 0.5223/0.6162 & 0.5132/0.6055 & 0.5185/0.6117 \\ 
2.0000 & 0.8175/0.9587 & 0.5323/0.6242 & 0.5251/0.6157 & 0.4808/0.5638 & 0.5520/0.6473 & 0.5326/0.6245 & 0.5436/0.6374 \\ 
2.5249 & 0.8432/0.9856 & 0.5492/0.6420 & 0.5351/0.6256 & 0.4953/0.5790 & 0.5684/0.6645 & 0.5431/0.6349 & 0.5572/0.6513 \\ 
2.7300 & 0.8348/0.9777 & 0.5438/0.6368 & 0.5274/0.6176 & 0.4902/0.5740 & 0.5624/0.6587 & 0.5353/0.6269 & 0.5503/0.6444 \\
\hline
\end{tabular}
}
\end{table*}

\begin{table}[!htbp]
\centering
\caption{\label{tab:fit_output}The fitted $z$-series expansion coefficients ($\alpha_0, \alpha_1$ for LB/UB) and the goodness-of-fit ($\chi^2/\text{d.o.f}$, $p$-value) for each form factor (FF).   }
\renewcommand{\arraystretch}{1.3}
\setlength{\tabcolsep}{4pt}
\resizebox{\columnwidth}{!}{
\begin{tabular}{c|c|c|c|c}
\hline
\multirow{2}{*}{FF}   &$\alpha_0$ &$\alpha_1$   &\multirow{2}{*}{$\frac{\chi^2}{\text{d.o.f.}}$} &\multirow{2}{*}{$p$-value} \\
     & (LB/UB) & (LB/UB)   &  &  \\
\hline
    $V$                  & 0.8398/0.9697 & $-6.3512$/$-5.9655$   & 0.0099                & 0.9998    \\ \hline
    $A_0$                & 0.5567/0.6431 & $-4.7812$/$-4.6402$   & 0.0099                & 0.9998    \\ \hline
    $A_1$                & 0.5094/0.5874 & $-0.9654$/$-0.1499$   & 0.0100                & 0.9998    \\ \hline
    $A_2$                & 0.4980/0.5751 & $-3.8532$/$-3.6418$   & 0.0098                & 0.9998    \\ \hline
    $T$                  & 0.5617/0.6485 & $-3.7889$/$-3.4385$   & 0.0099                & 0.9998    \\ \hline
    $T_1'$               & 0.5186/0.5982 & $-1.1648$/$-0.3711$   & 0.0100                & 0.9998    \\ \hline
    $T_2'$               & 0.5476/0.6320 & $-2.9184$/$-2.4206$   & 0.0101                & 0.9998    \\
\hline
\end{tabular}
}
\end{table}

In the actual fitting procedure,
we have restricted the NRQCD + lattice
inputs used for the $z$-series fits to $0 \le q^2 \le 3\,\mathrm{GeV}^2$, where the NRQCD
calculation is considered reliable.
Specifically, we first use Eq.~\eqref{nrlattformu} to generate the NRQCD + lattice synthetic data
$F_{\rm data}^{\mathrm{LB/UB}}(q_k^2)$ (derived from the lower/upper bound of lattice
inputs) for each form factor at a set of six discrete $q^2$ points
$\{q_k^2\}$ within the NRQCD-valid range $(0 \le q^2 \le 3\,\mathrm{GeV}^2)$.
These data points $\{q_k^2,\, F_{\rm data}^{\mathrm{LB/UB}}(q_k^2)\}$ are then used as
inputs to determine the parameters $\alpha^{\mathrm{LB/UB}}_{0}$ and
$\alpha^{\mathrm{LB/UB}}_{1}$ in Eq.~\eqref{zseries} by using the \texttt{Mathematica}
function \texttt{FindFit}.
To capture the theoretical uncertainty arising from  the lattice input data, we clarify that
instead of  fitting a single central value curve, we perform two separate fits for each form factor: we use the  lower-bound (LB) and upper-bound (UB) values of the synthetic data (the NRQCD + lattice results) as  separate inputs
and the $z$-series fits are performed independently for the lower-bound and upper-bound datasets,
which yields the lower and upper boundaries of the uncertainty bands of the NRQCD + lattice + $z$-series predictions shown in Figs.~\ref{fig:Vq2}--\ref{fig:AT2q2}. 
The explicit numerical values of these synthetic NRQCD + lattice input data points along with their corresponding $q^2$ values are listed in Table~\ref{tab:fit_input}.
And the fitted results of the $z$-expansion coefficients $\alpha^{\mathrm{LB/UB}}_{0},\alpha^{\mathrm{LB/UB}}_{1}$, together with
the corresponding statistical assessment of the goodness of the fits are provided in Table~\ref{tab:fit_output}.

In the statistical indicators, $\chi^2$ is defined as $\chi^2 = \sum_{k=1}^6 \left(F_{\rm fit}(q_k^2)-F_{\rm data}(q_k^2)\right)^2/{\sigma_k^2}$~\cite{fitbook,ParticleDataGroup:2024cfk},
where $F_{\rm fit}(q_k^2)=(F_{\rm fit}^{\mathrm{LB}}(q_k^2)+F_{\rm fit}^{\mathrm{UB}}(q_k^2))/2$
denotes the central value of the form factor obtained from the $z$-series fit in Eq.~\eqref{zseries}, calculated as the average of the lower- and upper-bound fit results,
while $F_{\rm data}(q_k^2)=(F_{\rm data}^{\mathrm{LB}}(q_k^2)+F_{\rm data}^{\mathrm{UB}}(q_k^2))/2$ represents the central value of the synthetic NRQCD + lattice input data generated via Eq.~\eqref{nrlattformu}, taken as the midpoint of its lower- and upper-bound values,
and $\sigma_k=(F_{\rm data}^{\mathrm{UB}}(q_k^2)-F_{\rm data}^{\mathrm{LB}}(q_k^2))/2$ is the effective standard deviation
for each data point $F_{\rm data}(q_k^2)$,~\footnote{The full width between the lower and upper bounds of the synthetic input data
is conservatively interpreted as covering a $\pm 1\sigma$ range, corresponding to a confidence interval of approximately $68\%$
under a normal (Gaussian) distribution assumption.}
taken as the half-width of the input data band.
The number of degrees of freedom is $\text{d.o.f.} = 6 - 2 = 4$, corresponding to six data points and two fit parameters ($\alpha_{0}, \alpha_{1}$).
The $p$-value of the fit is then calculated in \texttt{Mathematica} as
$p\text{-value} = 1 - \texttt{CDF}[\texttt{ChiSquareDistribution}[\text{d.o.f.}], \chi^2]$.

%

As shown in Table~\ref{tab:fit_output}, all form factors exhibit very good statistical goodness of fit,
with small $\chi^2/\text{d.o.f.}$ ($\sim 0.01\ll 1$) and
large $p$-values ($\simeq 0.9998\approx 1$),~\footnote{Our fit results yield $\chi^2/\text{d.o.f} \sim 0.01$ with $p \simeq 0.9998$ under
the conservative $\pm 1\sigma$ uncertainty assignment.
We further verify that even under a much more stringent $\pm 5\sigma$ interpretation of the input bands (i.e., assuming the LB/UB band corresponds to $\pm 5\sigma$,
approximately a $99.99994\%$ confidence level),
such that
 $\sigma_k=(F_{\rm data}^{\mathrm{UB}}(q_k^2)-F_{\rm data}^{\mathrm{LB}}(q_k^2))/10$,
the resulting $p$-value remains as high as $91\%$ for all form factors,
suggesting good consistency and reliability of the $z$-series parametrization.}
 indicating that the linear $z$-series parametrization $(N=1)$ provides a statistically consistent description of the synthetic NRQCD + lattice input data
within the adopted uncertainty bands in the NRQCD-valid low-$q^2$ region ($0 \le q^2 \le 3~\mathrm{GeV}^2$).
No statistically significant tension between the fit functions and the input data is observed,
supporting the reliability of the fitted $z$-series coefficients and justifying their use in interpolating and extrapolating the form factors
across the full physical $q^2$ range for phenomenological predictions and applications.


Finally, the NRQCD + lattice + $z$-series predictions for the $B_c^*\to\eta_c$ form factors are shown in Figs.~\ref{fig:Vq2}--\ref{fig:AT2q2}, where the error bars reflect the uncertainties originating from the lattice QCD data errors.
We note that our NRQCD + lattice + $z$-series predictions for the $B_c^*\to\eta_c$ vector and axial-vector form factors $V(q^2)$ and $A_{0,1,2}(q^2)$, shown in Figs.~\ref{fig:Vq2}--\ref{fig:A2q2}, exhibit significant differences in their $q^2$ dependence compared to the corresponding LFQM results presented in Ref.~\cite{Wang:2024cyi}. For comparison, Figs.~\ref{fig:Vq2}--\ref{fig:A2q2} also include the LFQM predictions from Ref.~\cite{addnote} for these form factors.


\section{Phenomenological Applications}\label{PHENOMENOLOGICAL}

Based on the predicted form factors in Eq.~\eqref{zseries}, we can study their phenomenological applications to the decay widths and branching fractions for the semileptonic decays $B_c^*\to \eta_c+l\nu_l$.
From Ref.~\cite{Wang:2024cyi}, the differential decay widths of $B_c^*\to \eta_c+l\nu_l$ can be expressed in terms of form factors as follows:
\begin{align}
\frac{d\Gamma_L}{dq^2}=&\left(\frac{q^2-m_l^2}{q^2}\right)^2\frac{\sqrt{\lambda(q^2)} G_F^2 |V_{cb}|^2}{384m_{B_c^*}^3\pi^3 q^2}
  \Bigg\{
    3 m_l^2 \lambda(q^2) A_0^2(q^2)
    \nonumber\\&
    +{\bigg |}
 (m_{B_c^*}^2-m_{\eta_c}^2-q^2)(m_{B_c^*}+m_{\eta_c})
    A_1(q^2)
 \nonumber\\&
    -\frac{\lambda(q^2)}{m_{B_c^*}+m_{\eta_c}}A_2(q^2){\bigg |}^2 \frac{m_l^2+2q^2}{4m^2_{\eta_c}}
 \Bigg\},\\
\frac{d\Gamma_\pm}{dq^2}=&\left(\frac{q^2-m_l^2}{q^2}\right)^2\frac{ \sqrt{\lambda(q^2)} G_F^2 |V_{cb}|^2} {384m_{B_c^*}^3\pi^3}
 \Bigg\{ (m_l^2+2q^2)  \lambda(q^2)
   \nonumber\\&
\times
 {\bigg |}\frac{V(q^2)}{m_{B_c^*}+m_{\eta_c}}\mp
 \frac{(m_{B_c^*}+m_{\eta_c})A_1(q^2)}{\sqrt{\lambda(q^2)}}{\bigg |}^2
 \Bigg\},\\
\frac{d \Gamma_{T}}{d q^{2}}=&\,\frac{d \Gamma_{+}}{d q^{2}}+\frac{d \Gamma_{-}}{d q^{2}},
\\
\frac{d \Gamma}{d q^{2}}=&\,\frac{d \Gamma_{L}}{d q^{2}}+\frac{d \Gamma_{T}}{d q^{2}},
\end{align}
where $\lambda(q^2)=(m^{2}_{B_c^*}+m^{2}_{\eta_c}-q^{2})^{2}-4m^{2}_{B_c^*}m^{2}_{\eta_c}$, and $m_{l}$ is the mass of the lepton $l\in\{e,\mu, \tau\}$,
while $d\Gamma_L/dq^2$, $d\Gamma_{T}/dq^2$ and $d\Gamma/dq^2$ are respectively the longitudinal, transverse and total differential decay widths for the semileptonic decay $B_c^*\to \eta_c+l\nu_l$.
Then the semileptonic decay width can be calculated by the following formula~\cite{Cohen:2018vhw}:
\begin{align}
\Gamma(B_c^*\to\eta_c+ l\nu_l)=\int_{m_l^2}^{(m_{B_c^*}-m_{\eta_c})^2} \frac{d\Gamma}{dq^2} dq^2.
\end{align}
Furthermore, the branching fraction and the ratio of branching fractions can be defined as follows~\cite{Wang:2024cyi}:
\begin{align}
\mathcal{B}(B_c^*\to\eta_c+ l\nu_l)&=\frac{\Gamma(B_c^*\to\eta_c+ l\nu_l)}{\Gamma_{\mathrm{tot}}(B_c^*)},\\
\mathcal{R}_{\eta_c}&=\frac{\mathcal{B}(B^*_{c} \rightarrow \eta_c+  \tau {\nu}_{\tau})}{\mathcal{B}(B^*_{c} \rightarrow \eta_c+  e {\nu}_{e})}.
\end{align}


In the calculation,
the numerical values of the following physical quantities can be directly obtained from the PDG~\cite{ParticleDataGroup:2024cfk, Workman:2022ynf}:
\begin{align}\label{paravalues1}
& |V_{cb}|=0.0408, ~~~ G_F=1.16638 \times 10^{-5} \,\mathrm{GeV}^{-2},
	\nonumber\\
	&	m_e=0.510999 \,\mathrm{MeV}, ~~~m_{\mu}=0.10566 \,\mathrm{GeV},
\nonumber\\
&m_{\tau}=1.777 \,	\mathrm{GeV},~~~
	m_{\eta_c}=2.9841\, \mathrm{GeV}.
\end{align}
As the $B_c^*$ meson has not yet been experimentally detected, the numerical values of its mass and decay width are calculated using various theoretical approaches~\cite{Yang:2021crs, Mathur:2018epb, Gomez-Rocha:2016cji, Zhou:2017svh, Gregory:2009hq, Fulcher:1998ka, Ikhdair:2003ry, Martin-Gonzalez:2022qwd, Colquhoun:2015oha, Chaturvedi:2022pmn} and are therefore model dependent.
In this paper, the mass of the  $B_c^*$  meson is primarily taken from lattice QCD calculations~\cite{Mathur:2018epb, Gregory:2009hq},
while its decay width is mainly based on quark potential model results~\cite{Fulcher:1998ka, Martin-Gonzalez:2022qwd, Yang:2021crs}.
The specific numerical values are as follows:
\begin{align}\label{paravalues2}
&m_{B^*_c}= 6.331\, \mathrm{GeV}, ~~~	\Gamma_{\mathrm{tot}}(B_c^*)=60\,  \mathrm{eV}.
\end{align}

\begin{table*}[!htbp]
\begin{center}
	\caption{The theoretical predictions for the decay widths, branching fractions, and the ratio of branching fractions for $B_c^*\to\eta_c+l\nu_l$, where the uncertainties of our results come from  the lattice QCD data errors.}
	\label{tab:widthBr}
	\setlength{\tabcolsep}{2.2mm}
	\renewcommand{\arraystretch}{1.8}
		\begin{tabular}{c|c|c|c|c|c}
			\hline
		\multirow{2}{*}{}	
			& \multirow{2}{*}{$\text{This work}$}
            & \multicolumn{2}{c|}{$\text{LFQM}$}
			& \multirow{2}{*}{$\text{BS}$}
            & \multirow{2}{*}{$\text{QCDSR}$}
   \\
   \cline{3-4}
   &
   & \cite{addnote}
   & \cite{Wang:2024cyi}
   & \cite{Wang:2018ryc}
   & \cite{Wang:2012hu}
   \\
			\hline
			\multirow{1}*{$\frac{10^{14}}{\text{GeV}}\Gamma({B_c^*\rightarrow\eta_c+ e\nu_e})$}
			& \multirow{1}*{$2.541_{-0.446}^{+0.494}$}
& $4.20_{-1.79}^{+1.63}$
			& \multirow{1}*{$-$}
&$0.966^{+0.094}_{-0.084}$
& $0.686^{+0.225}_{-0.195}$
              \\
			{$\frac{10^{14}}{\text{GeV}}\Gamma({B_c^*\to\eta_c+ \mu\nu_{\mu}}) $}
			& {$2.529_{-0.442}^{+0.489}$}
& $4.18^{+1.73}_{-1.78}$
			& {$-$}
&$0.963^{+0.094}_{-0.084}$
& $0.684^{+0.224}_{-0.195}$
	            \\
			{$\frac{10^{14}}{\text{GeV}}\Gamma({B_c^*\to\eta_c+ \tau\nu_{\tau}}) $}
			& {$0.688_{-0.096}^{+0.104}$}
& $1.09_{-0.44}^{+0.35}$
			& {$-$}
&$0.290^{+0.029}_{-0.026}$
& $0.215^{+0.075}_{-0.065}$
		                \\
			\hline
			$10^7\mathcal{B}({B_c^*\to\eta_c+ e\nu_e})$
			& {$4.235_{-0.743}^{+0.824}$}
& $7.00^{+2.71}_{-2.98}$
			& {$4.48^{+1.14}_{-0.76}$}
& $4.20^{+0.41}_{-0.37}$
& $-$
		               \\
			{$10^7\mathcal{B}({B_c^*\to\eta_c+ \mu\nu_{\mu}})$}
			& {$4.214_{-0.736}^{+0.816}$}
& $6.96^{+2.89}_{-2.96}$
			& {$4.45^{+1.14}_{-0.75}$}
& $4.19^{+0.41}_{-0.37}$
&$-$
		               \\
			{$10^7\mathcal{B}({B_c^*\to\eta_c+ \tau\nu_{\tau}})$}
			& {$1.147_{-0.161}^{+0.173}$}
& $1.82^{+0.58}_{-0.73}$
			& {$1.03^{+0.26}_{-0.17} $}
& $1.26^{+0.13}_{-0.11}$
&$-$
		               \\
			\hline
			{$\mathcal{R}_{\eta_c}$}
			& {$0.271_{-0.076}^{+0.107}$}
& $0.26^{-0.01}_{+0.01}$
			& {$0.229^{+ 0.059}_{-0.059}$}
& $0.300$
&	$-$	                		               		     		    		
		   \\
			\hline
			\end{tabular}
\end{center}		
\end{table*}

With the above formulas and the values of the input parameters in hand,
we calculate the decay widths, branching fractions, and the ratio of branching fractions for the semileptonic decay processes $B_c^*\to\eta_c+l\nu_l$. The numerical results are presented in Table~\ref{tab:widthBr}, where the uncertainties arise from the errors in the lattice input data. For comparison, the table also includes the corresponding results obtained from other approaches such as the light-front quark model (LFQM)~\cite{Wang:2024cyi,addnote}, the Bethe-Salpeter (BS) method~\cite{Wang:2018ryc}, and QCD sum rules (QCDSR)~\cite{Wang:2012hu}.
As shown in the table, our calculated decay widths are significantly larger than those obtained using the BS and QCDSR methods. However, our results for the branching fractions and the ratio of branching fractions are in good agreement with those from the LFQM and BS approaches.

\section{Summary}\label{SUMMARY}

Within the NRQCD factorization framework, we calculate the NLO perturbative QCD corrections to the semileptonic decays $B_c^*\to\eta_c+l\nu_l$. We obtain the complete NLO analytical results of the vector, axial-vector, tensor, and axial-tensor form factors for the $B_c^*\to\eta_c$ transition, as well as their asymptotic expressions in the hierarchical heavy quark limit.

We then investigate the renormalization scale dependence of the LO results, asymptotic NLO results, and complete NLO results for the form factors at the maximum recoil point $q^2=0$. The numerical results show that the NLO corrections reduce the renormalization scale dependence compared to the LO results.
Furthermore, by studying the  $q^2$ dependence of the NLO corrections to the form factors, we find that in the low $q^2$ region the NLO corrections are numerically sizable but remain convergent, whereas in the high $q^2$ region, the perturbative convergence begins to break down, indicating limitations of NRQCD at large momentum transfer.

To determine the product $\Psi_{B_c^*}(0)\Psi_{\eta_c}(0)$ of the meson wave functions at the origin in form factors, we approximate $\Psi_{B_c^*}(0)$ by $\Psi_{B_c}(0)$ using heavy quark spin symmetry. Then, $\Psi_{B_c}(0)\Psi_{\eta_c}(0)$ can be extracted by combining the NLO NRQCD results for the $B_c\to\eta_c$ vector form factors with the corresponding lattice QCD results.
As a result, this allows us to provide the theoretical predictions for the $B_c^*\to\eta_c$ form factors in the low $q^2$ region. By applying the $z$-series extrapolation method, we further extend the predictions across the full physical $q^2$ range, resulting in the NRQCD + lattice + $z$-series predictions for the form factors.

Based on the form factor predictions, we finally calculate the decay widths and branching fractions of the semileptonic processes $B_c^*\to\eta_c+l\nu_l$ for different lepton flavors.
These computations not only provide precise theoretical predictions that can assist in the experimental  search for  the $B_c^*$ meson, but also serve as valuable inputs for testing the Standard Model and exploring possible new physics beyond it.

\section*{Acknowledgements}

We thank Ruilin Zhu, Junfeng Sun and Su-Ping Jin for many helpful discussions.
The work is supported by the National Natural Science Foundation of China (Grant No. 12275067),
the Science and Technology R$\&$D Program Joint Fund Project of Henan Province  (Grant No. 225200810030),
the Science and Technology Innovation Leading Talent Support Program of Henan Province  (Grant No. 254000510039),
the National Key R$\&$D Program of China (Grant No. 2023YFA1606000),
the China Postdoctoral Science Foundation (Certificate Number: 2025M783381),
and the Key Research Project  for Higher Education Institutions of Henan Province (Grant No. 23A140012).
The work is partially supported by NSFC under Grants No. 12322503 and No. 12335003.

\appendix

\section{Asymptotic NLO-to-LO ratios for $B_c^*\to\eta_c$ (axial-)vector and (axial-)tensor form factors}\label{appendix123}

\begin{widetext}

In the hierarchical heavy quark limit, the asymptotic results for the NLO-to-LO ratios of the form factors are given by
\begin{align}
\frac{V^\text{NLO}}{V^\text{LO}}=&
1+\frac{\alpha_s}{4 \pi } \bigg\{\left(\frac{11 C_A}{3}-\frac{2n_f}{3} \right) \ln \frac{2 s y^2}{x} -\frac{10n_f}{9}
  +\left(-\frac{2 \ln x}{3}+\frac{2 \ln s}{3}+\frac{2 \ln 2}{3}+\frac{10}{9}\right)n_b
  -\frac{\ln
   x}{2}-\frac{\ln s}{2}
  \nonumber\\&
   -2 \ln 2+\frac{\pi ^2}{6}
+C_A
   \bigg[
   \left(\frac{3}{2}-4 s\right) \text{Li}_2(1-2 s)+(2 s-1) \text{Li}_2(1-s)
 -\frac{\ln
   ^2 x}{4}  -\left(\frac{\ln s}{2}+\frac{3 \ln 2}{2}+\frac{1}{2}\right) \ln   x
        \nonumber\\&
   -s \ln ^2 s
   +\left(\frac{s}{1-2 s}-4 s \ln 2\right) \ln s
   -2 s \ln ^2 2+\frac{s \ln 2}{1-2 s}+\frac{1}{9} \left(67-3 \pi ^2 s\right)\bigg]
        \nonumber\\&
      +C_F     \bigg[
  (8 s-3) \text{Li}_2(1-2 s)+(4-4 s) \text{Li}_2(1-s)
 +\frac{5 \ln^2 x }{4}
+\left(\frac{5 \ln s}{2}+6 \ln 2-\frac{23}{4}\right) \ln x
       \nonumber\\&
   +\left(2 s+\frac{7}{4}\right) \ln ^2 s
+\left(\frac{1}{4} \left(\frac{2}{2
   s-1}+\frac{6}{s-1}-11\right)+(8 s+3) \ln 2\right) \ln s
   +\left(4 s+\frac{3}{2}\right) \ln ^2 2
          \nonumber\\&
   +\left(\frac{1}{4 s-2}+1\right) \ln 2 +\frac{1}{12} \pi ^2 (8 s+17)-\frac{69}{4}\bigg]\bigg\},
\end{align}

\begin{align}
\frac{A_0^\text{NLO}}{A_0^\text{LO}}=& \frac{V^\text{NLO}}{V^\text{LO}} +  \frac{\alpha_s }{4 \pi }\bigg\{C_A \bigg[(2 s-1) \text{Li}_2(1-2 s)+\left(\frac{1}{2}-s\right) \text{Li}_2(1-s)+\frac{1}{4} (2 s-1) \ln
   ^2 s
 \nonumber\\&
   +\left(\frac{1}{2 s-1}+(2 s-1) \ln 2\right) \ln s
   +\left(s-\frac{1}{2}\right) \ln ^2 2+\left(\frac{1}{2 s-1}-1\right) \ln 2+\frac{1}{12} \pi ^2 (2 s-1)\bigg]
    \nonumber\\&
   +C_F
   \bigg[(2-4 s) \text{Li}_2(1-2 s)+(2 s-1) \text{Li}_2(1-s)
 +\left(\frac{1}{2}-s\right) \ln ^2 s +\left(\frac{(17-15 s) s-5}{(1-2 s)^2 (s-1)}+(2-4 s) \ln 2\right) \ln s
    \nonumber\\&
   +(1-2 s) \ln
   ^2 2
   +\frac{(s (8 s-11)+4) \ln 2 }{(1-2 s)^2}  +\frac{3 (8 s-5)-\pi ^2 (1-2 s)^2}{12 s-6}\bigg]\bigg\},
\end{align}

\begin{align}
 \frac{A_1^\text{NLO}}{A_1^\text{LO}}=& 1+   \frac{\alpha_s }{4 \pi }\bigg\{\left(\frac{11 C_A}{3}-\frac{2n_f}{3} \right) \ln  \frac{2 s y^2}{x}
    -\frac{10n_f}{9}+ \left(-\frac{2 \ln x}{3}+\frac{2 \ln
    s }{3}+\frac{2 \ln 2}{3}+\frac{10}{9}\right)n_b-\frac{\ln x}{3}-\frac{\ln s}{3}   -\frac{2 \ln 2}{3}
    \nonumber\\&
 +C_A\bigg[
   \left(\frac{8 s}{3}-1\right) \text{Li}_2(1-2 s)+\left(\frac{2}{3}-\frac{4 s}{3}\right) \text{Li}_2(1-s)-\frac{\ln^2 x }{6} +\left(-\frac{\ln s }{3}-\frac{1}{3}-\frac{\ln
    2 }{3}\right) \ln x
 +\frac{1}{3} (2 s-1) \ln^2 s
     \nonumber\\&
+\left(\frac{2-2 s}{6
   s-3}+\frac{4}{3} (2 s-1) \ln 2 \right) \ln s  +\frac{2}{3} (2 s-1) \ln ^2 2 +\frac{(6-10 s) \ln 2 }{6 s-3}  +\frac{1}{9} \left(2 \pi ^2 (s-1)+73\right)\bigg]
       \nonumber\\&
   +C_F \bigg[\left(2-\frac{16 s}{3}\right)
   \text{Li}_2(1-2 s)+\frac{2}{3} (4 s+1) \text{Li}_2(1-s)
    +\frac{\ln ^2 x }{2}
   +\left(\ln s -\frac{35}{6}+\frac{10 \ln 2 }{3}\right) \ln x
        \nonumber\\&
+\frac{1}{6} (11-8 s) \ln ^2 s +\left(\frac{1}{6} \left(\frac{4}{s-1}+\frac{2}{1-2 s}-17\right)-\frac{16}{3} (s-1) \ln
    2 \right) \ln s
            \nonumber\\&
-\frac{8}{3} (s-1) \ln ^2 2     +\left(\frac{1}{3-6 s}+4\right) \ln 2   +\frac{1}{18} \left(\pi ^2 (17-8
   s)-309\right)\bigg]
\bigg\},
\end{align}

\begin{align}
  \frac{A_2^\text{NLO}}{A_2^\text{LO}}=&1+   \frac{\alpha_s}{4 \pi } \bigg\{\left(\frac{11 C_A}{3}-\frac{2n_f}{3} \right) \ln  \frac{2 s y^2}{x}
     -\frac{10n_f}{9} +\left(-\frac{2 \ln x}{3}+\frac{2 \ln   s}{3}+\frac{2 \ln 2}{3}+\frac{10}{9}\right)n_b -\ln x-\ln s
     -6 \ln 2
              \nonumber\\&
+\frac{2 \pi ^2}{3}  +C_A\bigg[(12 (1-2 s)
   s+1) \text{Li}_2(1-2 s)+(6 s (2 s-1)-2) \text{Li}_2(1-s)-\frac{\ln
   ^2 x }{2}+(-\ln s-1-5 \ln 2) \ln x
   \nonumber\\&
   +\left(-6 s^2+3 s-1\right) \ln ^2 s
   +\left(-4 \left(6 s^2-3 s+1\right) \ln 2+\frac{1}{1-2 s}-3\right)
   \ln s+(6 (1-2 s) s-2) \ln ^2 2
     \nonumber\\&
   +\left(12 s+\frac{1}{1-2 s}-11\right) \ln 2+\pi ^2 (1-2 s) s +\frac{49}{9}\bigg]
   +C_F\bigg[   (24 s (2 s-1)-2) \text{Li}_2(1-2 s)
        \nonumber\\&
+6 \left(-4 s^2+2 s+1\right) \text{Li}_2(1-s)+\frac{7 \ln ^2 x }{2}+\left(7 \ln s -\frac{11}{2}+14 \ln 2 \right) \ln x
+\left(6 s
   (2 s-1)+\frac{11}{2}\right) \ln ^2 s
   \nonumber\\&
   +\left(12 \left(4 s^2-2 s+1\right) \ln 2
   +\frac{s (4 (s-7) s+25)-5}{2 (1-2 s)^2 (s-1)}\right) \ln
    s
    +6 \left(4 s^2-2 s+1\right) \ln ^2 2
    \nonumber\\&
    +\frac{25-2 s \left(48 s^2-98 s+63\right) }{(1-2 s)^2} \ln 2
    +\frac{1}{2 s-1} +\pi ^2 \left(4 s^2-2s+\frac{25}{6}\right)-\frac{37}{2}\bigg]
\bigg\},
\end{align}

\begin{align}
  \frac{T^\text{NLO}}{T^\text{LO}}=&1+      \frac{\alpha_s}{4 \pi } \bigg\{\left(\frac{11 C_A}{3}-\frac{2n_f}{3} \right) \ln \frac{2 s y^2}{x} -\frac{10 n_f}{9}
   +\left(-\frac{2 \ln x}{3}+\frac{2 \ln s}{3}+\frac{2 \ln 2}{3}+\frac{10}{9}\right)n_b -\frac{2 s \ln x}{4
   s+1}-\frac{2 s \ln s}{4 s+1}
   \nonumber\\&
   +\left(\frac{4}{4 s+1}-2\right) \ln 2+\frac{\pi ^2 (4 s-2)}{24 s+6}
   +C_A \bigg[\left(-2
   s-\frac{5}{4 s+1}+3\right) \text{Li}_2(1-2 s)+\left(\frac{2-7 s}{4 s+1}+s\right) \text{Li}_2(1-s)
   \nonumber\\&
-\frac{s \ln ^2 x}{4 s+1}+\left(-\frac{2 s}{4 s+1}+\frac{(2-6 s) \ln 2}{4 s+1}-\frac{2 s \ln s}{4
   s+1}\right) \ln x
+\frac{\left(s-2 s^2\right) \ln ^2 s}{4 s+1}
+\left(\frac{4 (1-2 s) s \ln 2}{4 s+1}-\frac{2 s}{4 s+1}\right) \ln s
    \nonumber\\&
   +\frac{2 (1-2 s) s \ln ^2 2}{4 s+1}-\frac{6 s \ln 2}{4
   s+1}
         +\frac{3 \pi ^2 \left(-2 s^2+s-1\right)+268 s+85}{36
   s+9}
   \bigg]
   +C_F \bigg[\left(4 s+\frac{10}{4 s+1}-6\right) \text{Li}_2(1-2 s)
     \nonumber\\&
   +\frac{(4 (5-2 s) s-2) \text{Li}_2(1-s)}{4 s+1}
   +\frac{(5 s-1) \ln ^2 x}{4
   s+1}+\left(-\frac{23 s+6}{4 s+1}+\frac{2 (5 s-1) \ln s}{4 s+1}+\left(6-\frac{8}{4 s+1}\right) \ln 2\right) \ln x
      \nonumber\\&
   +\left(\frac{4 s}{4 s+1}+s\right) \ln ^2 s
   +\left(\frac{-22 s^2+s+7}{8 s^2-2
   s-1}+\left(4 s+\frac{2}{4 s+1}\right) \ln 2\right) \ln s
  +\left(2 s+\frac{1}{4 s+1}\right) \ln ^2 2
      \nonumber\\&
   +\frac{2 \left(12 s^2-6 s+1\right) \ln 2}{8 s^2-2 s-1}+\frac{\pi ^2 (s (4 s+15)-2)-3 (69
   s+17)}{12 s+3}\bigg]\bigg\},
\end{align}

\begin{align}
 \frac{{T'_1}^\text{NLO}}{{T'_1}^\text{LO}}=&1+
\frac{\alpha_s}{4 \pi } \bigg\{\left(\frac{11 C_A}{3}-\frac{2n_f}{3} \right) \ln \frac{2 s y^2}{x}
-\frac{10n_f}{9} + \left(-\frac{2 \ln
   x}{3}+\frac{2 \ln s}{3}+\frac{2 \ln 2}{3}+\frac{10}{9}\right)n_b+\frac{2 s \ln x}{1-6 s}
          \nonumber\\&
   +\frac{2 s \ln s}{1-6 s}
   +\frac{(4 s+2) \ln 2}{1-6
   s}+\frac{\pi ^2}{18 s-3}
+C_A
   \bigg[\frac{2 (s-1) (4 s-1) \text{Li}_2(1-2 s)}{6 s-1}+\frac{\left(4 s^2-6 s+2\right) \text{Li}_2(1-s)}{1-6 s}
          \nonumber\\&
   +\frac{s \ln ^2 x}{1-6 s}
   +\left(\frac{2 s}{1-6
   s}+\frac{2 s \ln s}{1-6 s}+\frac{(2 s+2) \ln 2}{1-6 s}\right) \ln x+\frac{s (2 s-3) \ln ^2 s}{6
   s-1}+\left(\frac{2 s}{1-6 s}+\frac{4 (2 s-3) s \ln 2}{6 s-1}\right) \ln s
          \nonumber\\&
   +\frac{2 s (2 s-3) \ln ^2 2}{6 s-1}+\frac{6 s \ln 2}{1-6 s}+\frac{438 s+3 \pi ^2 (s (2 s-5)+1)-85}{54 s-9}\bigg]+C_F
   \bigg[-\frac{4 (s-1) (4 s-1) \text{Li}_2(1-2 s)}{6 s-1}
     \nonumber\\&
   +\frac{\left(8 s^2+2\right) \text{Li}_2(1-s)}{6 s-1}+\frac{(3 s+1) \ln ^2 x}{6 s-1}+\left(\frac{6-35 s}{6
   s-1}+\left(\frac{3}{6 s-1}+1\right) \ln s+\frac{(20 s+2) \ln 2}{6 s-1}\right) \ln x
     \nonumber\\&
 +\frac{(13-4 s) s \ln
   ^2 s}{6 s-1}+\left(\frac{(27-34 s) s-7}{4 s (3 s-2)+1}+\frac{(8 (5-2 s) s-2) \ln 2}{6 s-1}\right) \ln s+\frac{(4 s (2 s-5)+1) \ln ^2 2}{1-6 s}
   \nonumber\\&
   +\frac{(16
   s (2 s-1)-2) \ln 2}{4 s (3 s-2)+1}  +\frac{-309 s+\pi ^2 ((19-4 s) s+2)+51}{18 s-3}\bigg]\bigg\},
\end{align}

\begin{align}
\frac{{T'_2}^\text{NLO}}{{T'_2}^\text{LO}}=&\frac{{T}^\text{NLO}}{{T}^\text{LO}}+\frac{\alpha_s }{4 \pi }\bigg\{C_A \bigg[-\frac{4 (s-1) s \ln s}{(2 s-1) (4 s+1)}-\frac{4 (s-1) s \ln 2}{(2 s-1) (4 s+1)}\bigg]
+C_F \bigg[-\frac{4 (3 s-2) (4 s-1) s \ln s}{(1-2 s)^2 (4 s+1)}
  \nonumber\\&
-\frac{4 (s-1) s \ln 2}{(1-2 s)^2 (4 s+1)}
+\frac{4 (s-1)
   s}{(2 s-1) (4 s+1)}\bigg]\bigg\},
\end{align}
where $n_f=n_b+n_c+n_l$.

At the maximum recoil point ($q^2=0$ or $s=1$), the above expressions reduce to

\begin{align}\label{asyeqexp1}
\frac{{V}^\text{NLO}}{{V}^\text{LO}}=&1+
\frac{\alpha_s}{4 \pi }\bigg\{ \left(\frac{11 C_A}{3}-\frac{2 n_f}{3} \right)\ln \frac{2 y^2}{x}
-\frac{10n_f}{9}
+ \left(-\frac{2 \ln x}{3}+\frac{10}{9}+\frac{2 \ln 2}{3}\right)n_b-\frac{\ln x}{2}+\frac{\pi ^2}{6}-2 \ln   2
 \nonumber\\&
+C_A \bigg[-\frac{1}{4}
   \ln ^2 x+\left(-\frac{1}{2}-\frac{3 \ln 2}{2}\right) \ln x-2 \ln ^2 2-\ln 2-\frac{\pi ^2}{8}+\frac{67}{9}\bigg]
   +C_F \bigg[\frac{5 \ln
   ^2 x}{4}+\left(6 \ln 2-\frac{23}{4}\right) \ln x
   \nonumber\\&
   +\frac{11 \ln ^2 2}{2}+\frac{3 \ln 2}{2}+\frac{5 \pi ^2}{3}-\frac{63}{4}\bigg]\bigg\},
\end{align}

\begin{align}\label{asyeqexp2}
\frac{A_0^\text{NLO}}{A_0^\text{LO}}=&\frac{{V}^\text{NLO}}{{V}^\text{LO}}+
\frac{\alpha_s }{4 \pi }\bigg\{\frac{1}{2} C_A \ln ^2 2+C_F \left(-\frac{3}{2}-\ln ^2 2+\ln 2\right)\bigg\},
\end{align}

\begin{align}\label{asyeqexp3}
\frac{A_1^\text{NLO}}{A_1^\text{LO}}=&1+
\frac{\alpha_s }{4
   \pi }\bigg\{ \left(\frac{11 C_A}{3}-\frac{2n_f}{3} \right)\ln  \frac{2 y^2}{x}
    -\frac{10n_f}{9}
  +\left(-\frac{2 \ln x}{3}+\frac{10}{9}+\frac{2 \ln 2}{3}\right)n_b-\frac{\ln x}{3}-\frac{2 \ln 2}{3}
     \nonumber\\&
   +C_A \bigg[-\frac{1}{6}
   \ln ^2 x +\left(-\frac{1}{3}-\frac{\ln 2 }{3}\right) \ln x +\frac{2 \ln^2 2 }{3}-\frac{4 \ln 2 }{3}-\frac{5 \pi ^2}{36}+\frac{73}{9}\bigg]
      \nonumber\\&
   +C_F
   \bigg[\frac{\ln ^2 x }{2}+\left(\frac{10 \ln 2 }{3}-\frac{35}{6}\right) \ln x +\frac{11 \ln 2 }{3}+\frac{7 \pi ^2}{9}-\frac{33}{2}\bigg]
\bigg\},
\end{align}

\begin{align}\label{asyeqexp4}
\frac{A_2^\text{NLO}}{A_2^\text{LO}}=\frac{4A_0^\text{NLO}}{A_0^\text{LO}}-\frac{3A_1^\text{NLO}}{A_1^\text{LO}},
\end{align}

\begin{align}\label{asyeqexp5}
\frac{T^\text{NLO}}{T^\text{LO}}=\frac{{T'_1}^\text{NLO}}{{T'_1}^\text{LO}}=\frac{{T'_2}^\text{NLO}}{{T'_2}^\text{LO}}=\frac{2V^\text{NLO}}{5V^\text{LO}}+\frac{3A_1^\text{NLO}}{5A_1^\text{LO}}-\frac{\alpha_s C_F}{4\pi}.
\end{align}

\end{widetext}

\end{document}